%% file: main.tex
\documentclass[12pt]{article}
\usepackage{graphicx} 
\usepackage{hyperref}
\usepackage{amsmath}
\usepackage{cleveref}
\usepackage[version=4,arrows=pgf-filled,
textfontname=sffamily,
mathfontname=mathsf]{mhchem}
\usepackage{setspace}
\usepackage{caption}
\newcommand{\parallelsum}{\mathbin{\!/\mkern-5mu/\!}}
\usepackage[letterpaper, margin=1in]{geometry}
\title{Compact Circuit Models for Nanoantenna-Based Lightwave Electronics}
\author{Adina Bechhofer, Felix Ritzkowsky, Karl K. Berggren,\\ Phillip D. Keathley}
\date{October, 2026}

\begin{document}

\maketitle
\section*{Abstract}
As the field of nanoscale lightwave electronics matures from centering around fundamental science to focusing on device engineering, there is a growing need for compact models that can provide rapid and scalable quantitative analysis of the performance of petahertz-frequency optical-field-driven electronics.  In this work, we developed a circuit model framework for describing nano-scale lightwave electronics. Our framework captures the physics governing nanoscale optical devices, such as electrically-connected nanoantennas, using compact circuit models that predict their electromagnetic response, their electron emission, and their charge transfer dynamics. This approach allows scalable simulation of large integrated systems containing networks of nanoscale lightwave electronic components, which is not possible using full-wave electromagnetic and particle-in-cell simulation methods. We implemented the compact model in an free commercial circuit solver (LTspice) and validated its electromagnetic response against a full-wave electromagnetic solver (MEEP). We propose an experiment to fully benchmark the model's ability to capture non-linear charge-transfer-based coupling between devices. The circuit model implementation speeds up the electromagnetic analysis from hours to seconds and accounts for charge-driven coupling, enabling rapid quantitative studies of device operation which provide new insights into how femtosecond signals propagate through nanoscale lightwave electronic structures. We anticipate that the methods we introduce here will become essential to the development and performance analysis of nanoscale lightwave electronics for communication, computation, and signal processing at optical frequencies.   

\section{Introduction}
\input{introduction}
\section{Model}
\input{model}

\input{Application_example}
\section{Conclusion and outlook}
\input{conclusions}
\section{Acknowledgments}
This material is based upon work supported by the Department of the Navy, Office of Naval Research under Grant No. N000142412350 which supported the work related to modeling interconnect and analyzing their effects on arrays of structures. This material is based upon work supported by the National Science Foundation under Grant No. 2238575 which supported the work related to simulated demonstrations of ultrafast sampling with cross correlation applied to the compact model. This material is based upon work supported by the Air Force Office of Scientific Research under award number FA9550-18-1-0436 which supported the initial phases of the unit cell model. We acknowledge helpful discussions with Prof. Luca Daniel, Dr. John Simonaitis, and Dr. Shruti Nirantar. We thank Malick  Sere, Dr. Gian Luca Dolso, and Dr. Iliya Prigozhin for their helpful comments on the manuscript.  

\bibliographystyle{IEEEtran}
\linespread{1}
\bibliography{ref}
\newpage 
\appendix
\section{Appendix} 
\input{appendix}
\end{document}

%% file: introduction.tex
 The term `lightwave electronics' describes optically-driven systems with response times that are on the timescale of a cycle of the driving waveform or faster. Since optical fields have petahertz frequencies, those systems are also referred to as petahertz electronics. The Keldysh adiabatic parameter, $\gamma_k = \omega/\omega_t$, compares the rate of electron emission due to multi-photon absorption with the rate of electrons tunneling through a potential barrier. The parameter depends on the energy of the incident photons as well as the peak electric field of the pulse, such that $\gamma_k = \frac{\omega\sqrt{m_e \phi}}{eE_0}$, where $\omega$ is the photon frequency, $m_e$ is the effective electron mass, $\phi$ is the work function, $e$ is the elementary charge, and $E_0$ is the peak field of the incident pulse \cite{PhysRevA.104.023103, RevModPhys.90.021002}. Systems described by $\gamma_k >1$ belong to the photonic regime, where the dominant electron emission mechanism is driven by multi-photon absorption which depends on the energy of the absorbed from photons over many optical field cycles. In systems described by $\gamma_k < 1$, the peak fields bend the potential barrier between metal and vacuum sufficiently to cause electron emission to be dominated by tunneling through the barrier. The timescale of tunneling is sub-optical field cycle, meaning that systems with $\gamma_k < 1$ belong to the electronic regime \cite{heide_petahertz_2024}. 

Light wave pulses with carrier envelopes that last only for a few optical cycles can have peak fields on the order of GV/m, while delivering insufficient energy per pulse to cause damage to the systems \cite{yang_light_2020}. Ultrashort light pulses have been shown to induce petahertz ($10^{15}$ Hz) scale electron dynamics in gasses \cite{corkum_attosecond_2007}, solids \cite{ghimire_observation_2011, schultze_attosecond_2014}, and at material interfaces \cite{hommelhoff_field_2006, hommelhoff_ultrafast_2006, arai_all-solid-state_2023}. A few-cycle laser with central wavelength in the near-infrared was used to induce field-driven tunneling ionization currents from metallic needle tips \cite{hommelhoff_field_2006, kruger_attosecond_2011, herink_field-driven_2012}. Shortly there after, sub-cycle electron tunneling with ultrashort pulses was demonstrated from metallic nanostructures on a substrate \cite{dombi_ultrafast_2013}. In response to incident optical fields, nanostructures support surface plasmon modes, which enhance the driving surface field \cite{li_surface-plasmon_2013, hobbs_high-yield_2014} leading to higher electron emission with lower incident fields. Advancements in nanofabrication enable precise and reliable fabrication of structures with tips having nanometer-scale radii of curvature and gap sizes on the scale of few to tens of nanometers. Those geometrical features introduce an additional geometric field enhancement \cite{putnam_optical-field-controlled_2017} which, along with the plasmonic enhancement, results in orders of magnitude reduction in the peak power of the laser required to induce tunneling ionization \cite{zimmermann_toward_2019, ludwig_sub-femtosecond_2020, heide_petahertz_2024}.   

The electromagnetic and charge emission responses of single structures when driven by optical inputs enabled time resolution measurement of attosecond electric fields \cite{bionta_-chip_2021, luo_synthesis_2023} and detection of carrier envelop phase (CEP) shifts in driving pulses \cite{yang_light_2020} in arrays of nanostructures. A graphene-gold interface structure has been used to demonstrate petahertz logic on incident light pulses, leveraging the CEP-dependent current responses of graphene and graphene-gold interfaces \cite{boolakee_light-field_2022}. However, in order to extend the framework of lightwave electronic circuits into the realm of petahertz computing and communication with multi-step logic operations, memory, and data transfer, we must leverage the nonlinear coupling and charge transfer responses between electrically-connected structures. 

The bottleneck for realizing lightwave controlled petahertz devices that leverage nonlinear coupling and charge transfer between structures is the lack of scalable simulations to enable the design of such systems. 
To addresses the bottleneck in lightwave device design, we present a lightweight, few-parameter model that captures the most dominant physical effects of an ultrafast optically driven nano-emitter while lending itself to fast simulations. We chose circuit models as the paradigm because: (1) they are ubiquitous in resonant dynamics modeling \cite{burke_rf_2003}; (2) they are intuitive and physically meaningful; and (3) they naturally fit into the integrated circuit design workflow. 

Previous efforts have used circuit models and have leveraged antenna modeling to describe the electromagnetic response of nanoplasmonic structures to incident optical frequency fields \cite{engheta_circuit_2005, alu_wireless_2010, zhu_circuit_2014, eggleston_optical_2015, alu_theory_2013, staffaroni_circuit_2012}. Those models were constructed from first principles such as quasi-static treatments of plasmonic interfaces \cite{engheta_circuit_2005}, energy exchange between electric and magnetic fields \cite{zhu_circuit_2014}, or antenna radiation theory \cite{alu_wireless_2010, eggleston_optical_2015}. The focus of those previous works was almost entirely optical; predicting plasmon resonance frequencies, quality factors of far field resonance responses, and far field optical radiation patterns. With respect to implementation of lightwave electronics, the prior works fall short in three key areas. (1) They do not predict the local time dependent electric field distribution on a structure which is essential for modeling oriented towards charge-emitting devices. Additionally,  (2) they do not account for time delay and interferences due to nanowire interconnects. While there are works modeling propagation dynamics of plasmons on interconnects \cite{staffaroni_circuit_2012}, they do not account for coupling to additional plasmonic structures. Finally, (3) the previous works do not account for field emission from nanostructures or the charge transfer dynamics through the systems.   

In this work, we introduce a model that accurately describes the local electric field and electronic response to an incident optical field on various parts of a nanostructure. The model also accounts for coupling between nanostructures and electrical interconnects as well as field driven charge emission from the structure and subsequent charge transfer effects that have been ignored by prior work. 

\begin{figure}[htbp]
    \centering
    \includegraphics[width=0.98\linewidth]{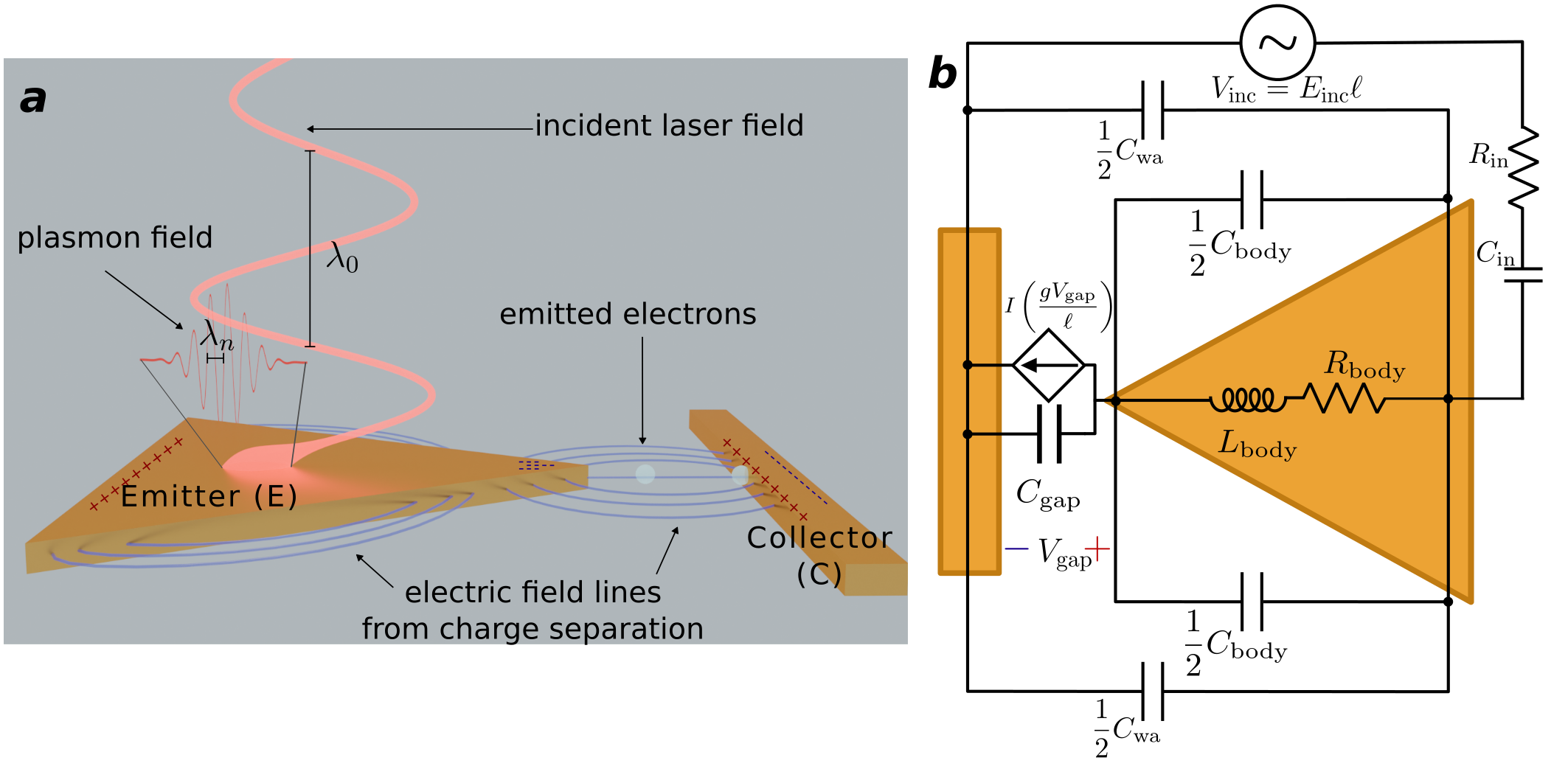}
    \caption{Compact circuit models are used to describe the electromagnetic response and the charge emission in field-driven charge-emitting nanostructures. \textbf{a} Individual triangular nanoantenna depicted with an incident optical field which induces charge separation and local electric fields on the surface of the structure. The variable $\lambda_0$ is the vacuum wavelength of the incident field and $\lambda_\text{n}$ is the wavelength of the induced surface plasmon. Electrons are tunnel-emitted from the tip of the antenna due to field emission. \textbf{b} Compact circuit model corresponding to a singular nanoantenna in an optical field. The incident pulse is represented by the voltage source with waveform proportional to the incident pulse field $E_\text{inc}$ such that $V_\text{inc}=E_\text{inc}\ell$, where $\ell$ is a length scale related to the gap distance. The incident voltage is coupled into the antenna through input impedance $R_\text{in}+C_\text{in}$. The optical resistance and optical inductance of the antenna body are given by $R_\text{body}$ and $L_\text{body}$. The self-capacitance of the antenna is given by $C_\text{body}$. The electric fields in the gap between the emitter's tip and the collector are accounted for by the gap capacitance $C_\text{gap}$, and the fields between the collector and the emitter's base is accounted for by $C_\text{wa}$. The voltage across the gap, $V_\text{gap}$ is proportional to the enhanced field at the tip of the antenna such that $V_\text{gap} = E_\text{tip}\ell/g$, where $g$ is a geometrical field enhancement. The electron emission from the emitter is described with a voltage-controlled current source.}
    \label{fig:big_pic}
\end{figure}

%% file: model.tex
We consider a general two part asymmetrical nano structure in an optical field of a few-cycle femtosecond laser pulse. Asymmetric triangular gold nanoantennas with nanometer scale gaps are of interest because they have nonlinear current rectifying characteristics \cite{putnam_optical-field-controlled_2017}, high field enhancements, broadband wavelength resonance responses, and polarization sensitivity \cite{hobbs_mapping_2017}, which are valuable for implementation of PHz lightwave devices. More importantly, asymmetric triangular nanoantennas have been fabricated and tested by several research groups, experimentally showing optical-field-driven tunneling emission \cite{putnam_optical-field-controlled_2017, yang_light_2020,bionta_-chip_2021, turchetti_electron_2022}. The experimental implementation makes the structure perfect for benchmarking with observable data. 

The conceptual picture in \cref{fig:big_pic}.\textbf{a} shows an incident external field inducing a plasmonic mode confined to the surface of a metallic nanostructure. The wavelength of the plasmonic mode is shorter than that of the incident field in vacuum by a factor of $\sim 2.7$ \cite{weeber_near-field_2001}. The oscillating plasmon generates oscillating charges, developing optical kinetic inductance and resistance. Instantaneously, opposite charges accumulate at opposite ends of the structure, creating a potential difference with electric field lines that stretch between the tip of the triangle and its base, depicted with blue lines in \cref{fig:big_pic}.\textbf{a}. This potential difference can be attributed to structure's self capacitance. The charges on the structure's tip also establish a potential difference across the gap between the tip of the antenna and the thin wire, corresponding to the capacitance of the gap. An oscillating potential develops across the wire and the base of the triangle, indicated by another capacitance. At the tip of the triangle, where the field is at its highest due to geometric enhancement, electrons are tunnel-emitted from the surface of the metal and are swept across the gap, where they are collected by the wire, as shown in \cref{fig:big_pic}.\textbf{a}. 

\Cref{fig:big_pic}.\textbf{b} shows the circuit model of an asymmetric two-part structure shown in \textbf{a}. The model is compact with just three nodes: one for the base of the emitter, one for the tip of it, and one for the collector. Following the convention from RF antenna modeling, the incident optical field is modeled as a scaled voltage source, $V_{\text{inc}} = E_\text{inc} \ell$, where $E_\text{inc}$ is the incident optical field profile and $\ell$ is a length scaling factor related to the gap length. The incident voltage source stretches across the structure and is coupled into the antenna emitter through a resistor, $R_{\text{in}}$, to account for radiative loss \cite{alu_theory_2013,ellingson_electromagnetics_2020, ellingson_radio_2023}. A capacitive input impedance $C_{\text{in}}$, is added in series to account for the phase shifts between the vacuum field and the plasmonic mode. The optical kinetic inductance and optical resistance of the body of the antenna are described by $L_{\text{body}}$ and $R_{\text{body}}$ along the length of the antenna. The antenna's self capacitance, $C_\text{body}$ which accounts for the potential difference between the antenna tip and its base, is connected in parallel to $R_{\text{body}}$+$L_{\text{body}}$. The potential difference across the gap is accounted for by $C_\text{gap}$, and the capacitance between the collector wire and the triangular structure's base is represented by $C_\text{wa}$. 
The charge emission rate can be described by a non-linear rectifying current source, $I(V_{\text{gap}})$, that is controlled by the gap voltage. The gap voltage, $V_\text{gap}$, is proportional to the field at the tip of the antenna, such that $V_\text{gap} = \frac{E_\text{tip}\ell}{g}$ where $g$ is e geometrical enhancement factor and $\ell$ is the aforementioned length scale factor. The model can be extended easily to describe a symmetric antenna by reflecting the sub-circuit corresponding to the emitter component across the gap. The current source accounting for charge emission would be non-rectifying for the symmetric structure. The selection of specific values for the circuit parameters of the compact model defines the peak frequency and the bandwidth of the structure's resonance, corresponding to the material and geometry of the structure. 

The propagation dynamics of optical modes on interconnects account for time delay in signal propagation, cross-talk between antennas, and reflections which can affect resonances of antenna elements. In line with prior work \cite{staffaroni_circuit_2012}, we model the plasmonic interconnects as lossy transmission lines. For conciseness, the interconnects are not shown in \cref{fig:big_pic}, but are discussed in detail in \Cref{sec:verification_and_analysis} where the coupling between an antenna and interconnects is introduced and analyzed in detail. 

\section{Verification and Analysis}
\label{sec:verification_and_analysis}
The analysis and verification of the model are done in two phases: in phase I, the electromagnetic characteristics of individual isolated structures and multipart coupled structures are evaluated. In phase II, non-linear charge emission and transport characteristics are added to the model. The effects of the non-linear charge transfer dynamics are analyzed in a simulated correlation measurement which provides direct one-to-one comparison with experiment anmd can serve to benchmark the model.   

\subsection*{Phase I: Verification of Electromagnetic Dynamics in Stand-alone and Coupled Structures}
\begin{figure}[htbp]
    \centering
    \includegraphics[width=0.9\linewidth]{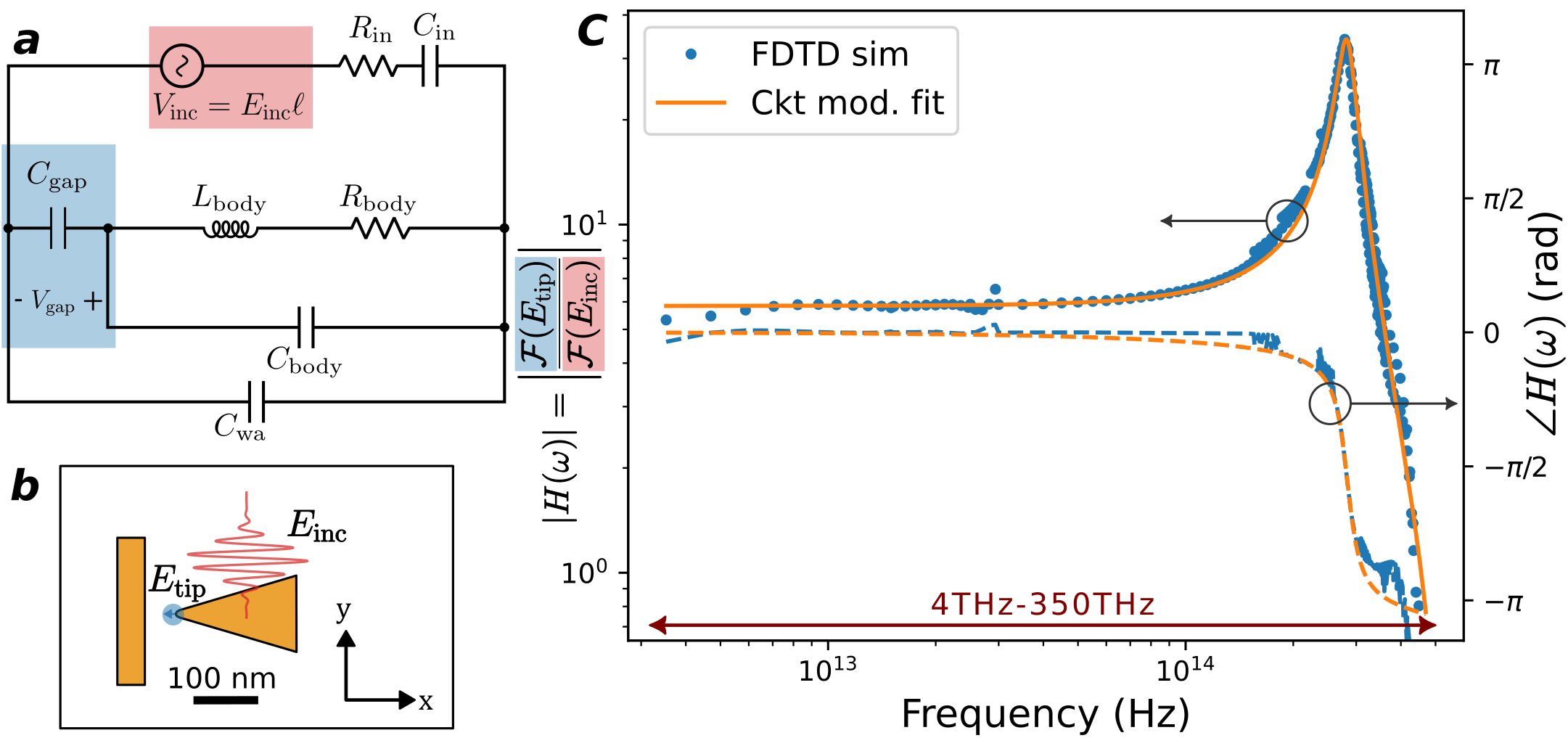}
    \caption{Least of squares fitting of circuit model parameters to finite difference time domain simulation results of 200 nm antenna. The circuit model describes FDTD results accurately over the frequency range of 5 THz-350 THz. \textbf{a} Compact circuit model of the nanoantenna-collector pair.  $H_\text{ckt}(\omega) = \frac{V_{\text{gap}}(\omega)}{V_{\text{opt}}(\omega)}$, where $V_{\text{gap}}(\omega)$ is the voltage across the gap capacitor (highlighted in blue) and $V_{\text{opt}}(\omega)$ is the input voltage representing the optical field (highlighted in red). \textbf{b} The simulated structure consists of a triangle of height of 200 nm, base of 100 nm, and a rounded tip with a 10 nm radius of curvature. The collector was a 50-nm-wide wire, separated from the triangular tip by a 40 nm gap distance. The material for structure was defined as a thin gold layer of 20 nm thickness sitting on an $\text{SiO}_2$ substrate. The incident optical field was X-polarized. The transfer function is defined as the Fourier transform of the simulated field at the tip divided by the Fourier transform of the incident optical field. \textbf{c} Comparison of transfer function calculated from a set of FDTD simulations of structure in the inset and the analytical transfer function of the circuit in \textbf{a}. The circuit parameters in \textbf{a} were obtained by tuning the initial guess values with a least of squares fitting of the transfer function. Resulting model agrees with $R^2=0.957$ over a 200 THz range. }
    \label{fig:unitcell-fitting}
\end{figure}

For the analysis of the electromagnetic aspects of the model, we removed the non-linear charge emission source from the circuit model. The reduced circuit is shown in \cref{fig:unitcell-fitting}.\textbf{a}. The transfer function of the circuit model is defined as follows: 
$H_\text{ckt}(\omega) = \frac{V_{\text{gap}}(\omega)}{V_{\text{opt}}(\omega)}$, where $V_{\text{gap}}(\omega) \propto E_\text{tip}(\omega)$ is the voltage across the gap proportional to the field at the antenna tip and $V_{\text{opt}}(\omega) \propto E_\text{inc}(\omega)$ is the input voltage proportional to the incident optical field. With $Z_\text{body} = (R_\text{body} + j\omega L_\text{body})\parallelsum \frac{-j}{\omega C_\text{body}}$, the transfer function can be written as:
\begin{equation}
    H_\text{ckt}(\omega) = \frac{\frac{-j}{\omega C_\text{gap}}}{Z_\text{body} + \frac{-j}{\omega C_{\text{gap}}}} \frac{\frac{-j}{\omega C_{\text{wa}}}\parallelsum \left(Z_\text{body} +\frac{-j}{\omega C_\text{gap}}\right)}{\frac{-j}{\omega C_{\text{wa}}}\parallelsum \left(Z_\text{body} +\frac{-j}{\omega C_\text{gap}}\right) + R_{\text{in}} + \frac{-j}{\omega C_{\text{in}}}}.
\end{equation}
 The full derivation of the analytical transfer function can be found in \cref{sec:analytical-transfxn-deriv}.

MEEP \cite{OSKOOI2010687, alma9935236045106761} was used to perform a set of full-wave finite difference time domain (FDTD) simulations of a 20 nm thick gold structure comprising of a triangular antenna of height 200 nm and a wire collector of width 50 nm. The structure was illuminated from above by a short envelope pulse polarized along the axis of the height of the triangular antenna. The simulated structure is depicted in \cref{fig:unitcell-fitting}.\textbf{b}. The frequency dependent transfer function was calculated from the simulation results as follows $H_\text{FDTD}(\omega) = \frac{\mathcal{F}(E_{\text{tip}})(\omega)}{\mathcal{F}(E_{\text{inc}})(\omega)}$, where $\mathcal{F}(E_{\text{tip}})$ is the Fourier transform of the time dependent normal component of the electric field at the tip, and $\mathcal{F}(E_{\text{inc}})$ is the time dependent incident field. More information about simulation and post processing procedure can be found in \cref{sec:fdtd-sims-unit-cell}. The amplitude of the FDTD simulated transfer function is plotted in blue dots in \cref{fig:unitcell-fitting}.\textbf{c}, and its phase is plotted in blue dashes with the scale indicated on the right y-axis.

The initial inductance and resistance values of the body resistor and body inductor in the circuit model were calculated from the Drude model \cite{alma990017320150106761, zhu_circuit_2014} applied to a simplified rectangular structure of similar dimensions to the antenna. The calculation yielded $L_{\text{body}, i} = 0.12 $pH and $R_{\text{body}, i} = 4.4 \Omega$. The initial gap capacitance was estimated by a parallel plate approximations of the simplified structure with $C_{\text{gap},i} = 0.22$ aF. The derivation can be found in \cref{sec:initial-circuit-param-extraction}. The final circuit model parameter values were calculated from a least of squares optimization which minimized the root mean squared of the difference between the normalized analytical circuit transfer function, $H_{\text{ckt}}(\omega)/|H_{\text{ckt}}(\omega)|_{\text{max}}$ and the normalized antenna transfer function, $H_\text{FDTD}(\omega)/|H_\text{FDTD}(\omega)|_\text{max}$ calculated from the FDTD simulation, as detailed in \cref{sec:los-fitting}. The scaling factor that emerges from the fitting $g = \frac{|H_\text{FDTD}(\omega)|_\text{max}}{|H_\text{ckt}(\omega)|_\text{max}} = 7.22$ is consistent with values of the local geometrical field enhancement of the tip, often referred to as the lightning rod effect, reported in literature for similar structures \cite{turchetti_electron_2022}. The resulting amplitude and phase of the analytical circuit transfer function are shown with orange solid and broken curves respectively in \cref{fig:unitcell-fitting}.\textbf{b}. The curve of the analytical transfer function and the curve of the FDTD transfer function show agreement with an $R^2 = 0.957$. More details about the metric can be found in \cref{sec:los-fitting}. 

\begin{figure}[htbp]
    \centering
    \includegraphics[width=0.95\linewidth]{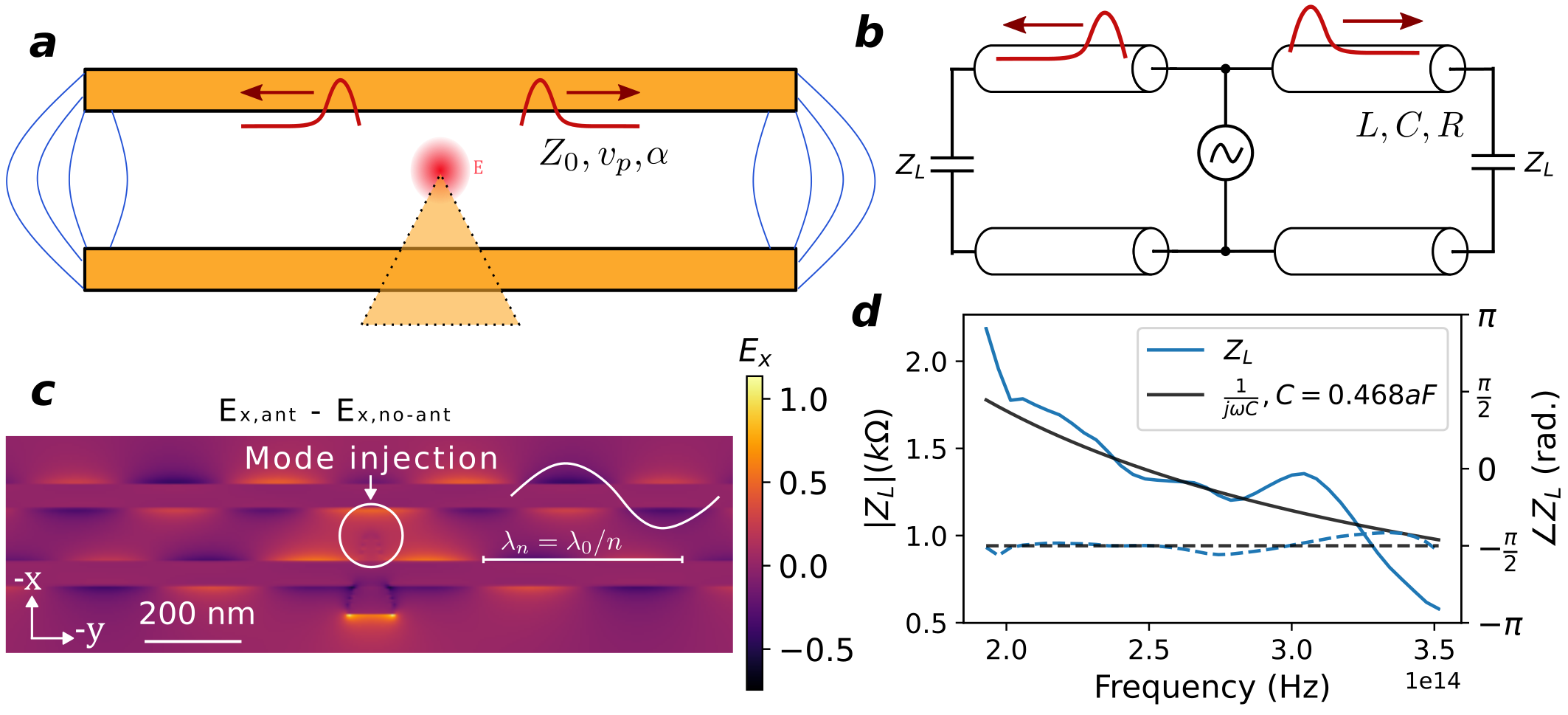}
    \caption{Nanowire interconnects act as transmission lines propagating plasmonic modes. \textbf{a} Schematic of simulation used to extract transmission line properties. By subtracting the results of two simulations, one with the antenna in place and one without, a local source was inserted to the center of the wire and the propagating mode was observed along the interconnects. See text for details. \textbf{b} Corresponding transmission line circuit schematic of the system simulated. \textbf{c} Time snapshot from an FDTD simulation describing the propagation of a plasmonic mode on the nanowire. The heatmap color corresponds to the value of the normal component of the electric field ($E_x$) on the wire. \textbf{d} The calculated load impedance at the end of the wires has capacitive characteristics. A capacitive load impedance corresponding to 0.468 aF (solid black curve) provides a reasonable fit to the calculated load impedance (solid blue curve).}
    \label{fig:transmission-line-char}
\end{figure}

Having established an accurate antenna model, we then extended the model by including electronic interconnects. The interconnects considered in this work fall outside the scope of the analytical model introduced by Staffaroni et al. \cite{staffaroni_circuit_2012} due to geometrical constraints and loss conditions. Here, we take an empirical approach to extracting transmission line parameters from a full wave simulation. To circumvent numerical instability associated with point sources, we implemented a two-part simulation with a plane wave input, as shown in \cref{fig:transmission-line-char}.\textbf{a}. In the first simulation, the structure consisted of a pair of  2.6 micron long, 50 nm wide, 20 nm thick, gold nanowires spaced 115 nm apart. A 200 nm height gold triangular antenna (shown in \cref{fig:transmission-line-char}.\textbf{a} with a dotted outline) was embedded in one wire with its tip pointing towards the other wire, forming a 40 nm gap. The structure in the second simulation consisted of the two wires alone (shown in \cref{fig:transmission-line-char}.\textbf{a} with solid outlines). Both simulations were driven by identical X-polarized plane wave sources with central wavelength in the near infrared (near-IR). The resulting electric field from the second simulation (wires alone) was subtracted from the electric field results of the first simulation (antenna + wire pair). The resulting time dependent electric field allows analysis of the local mode injected by the triangular antenna onto the surfaces of the nanowires in isolation. The corresponding transmission line circuit is shown in \cref{fig:transmission-line-char}.\textbf{b}, where the segments of the nanowire above and below the point of mode injection are treated as distinct transmission line segments in parallel. A time domain snapshot of the subtracted simulation is shown in \cref{fig:transmission-line-char}.\textbf{c}. The transmission line coefficients were extracted from the subtracted simulation using only the time steps before the arrival of reflections. The characteristic impedance of the wire was calculated as follows $Z_0 (\omega)= \frac{E(\omega)}{H(\omega)} = 1.65 \sqrt{\frac{\mu_0}{\epsilon_0}}$, where $E(\omega)$ is the frequency dependent electric field and $H(\omega)$ is the frequency dependent magnetic field, both sampled on the surface of the wire about 200 nm away from the point of mode injection. While $Z_0 (\omega)$ is frequency dependent, the calculated value is well approximated as a constant for the frequency range of interest. The surface-confined phase velocity, $v_p = 0.35 c$, was obtained from tracking mode propagation along the wire in time domain. The plasmonic loss $\alpha = 1.29 \times 10^{-4} \ \Omega \text{nm}^{-1}$ was observed from the decay of the mode along its travel.

The load impedance responsible for the reflections was calculated by isolating the forward and backwards propagating electric field modes on the wire. The forward propagating mode can be written as $E^{
+} = \frac{1}{2} (E(\omega) + Z_0 H(\omega))$ and the backwards propagating mode can be written as $E^{-} = \frac{1}{2} (E(\omega) - Z_0 H(\omega))$ evaluated at the end of the wire. The load impedance then can be written as $Z_L = Z_0\frac{E^+ + E^-}{E^+ - E^-}$. As shown in \cref{fig:transmission-line-char}.\textbf{d}, the load is strongly capacitive with impedance $Z_L = \frac{-j}{\omega 0.468 \text{aF}}$.

The process for solving for the distributed resistance per unit length, inductance per unit length, and capacitance per unit length, from the values of $Z_0$, $v_p$, and $\alpha$, is described in \cref{sec:app-tline-simulation} alongside additional simulation details. For the geometry simulated in \cref{fig:transmission-line-char}.\textbf{a}, the transmission line parameters are $L=5.928 \ \text{H}\text{nm}^{-1}$, $R=0.0804 \ \Omega\text{nm}^{-1}$, and $C=1.53\times 10^{-20} \ \text{F}\text{nm}^{-1}$. 

\begin{figure}[htbp]
    \centering
    \includegraphics[width=0.85\linewidth]{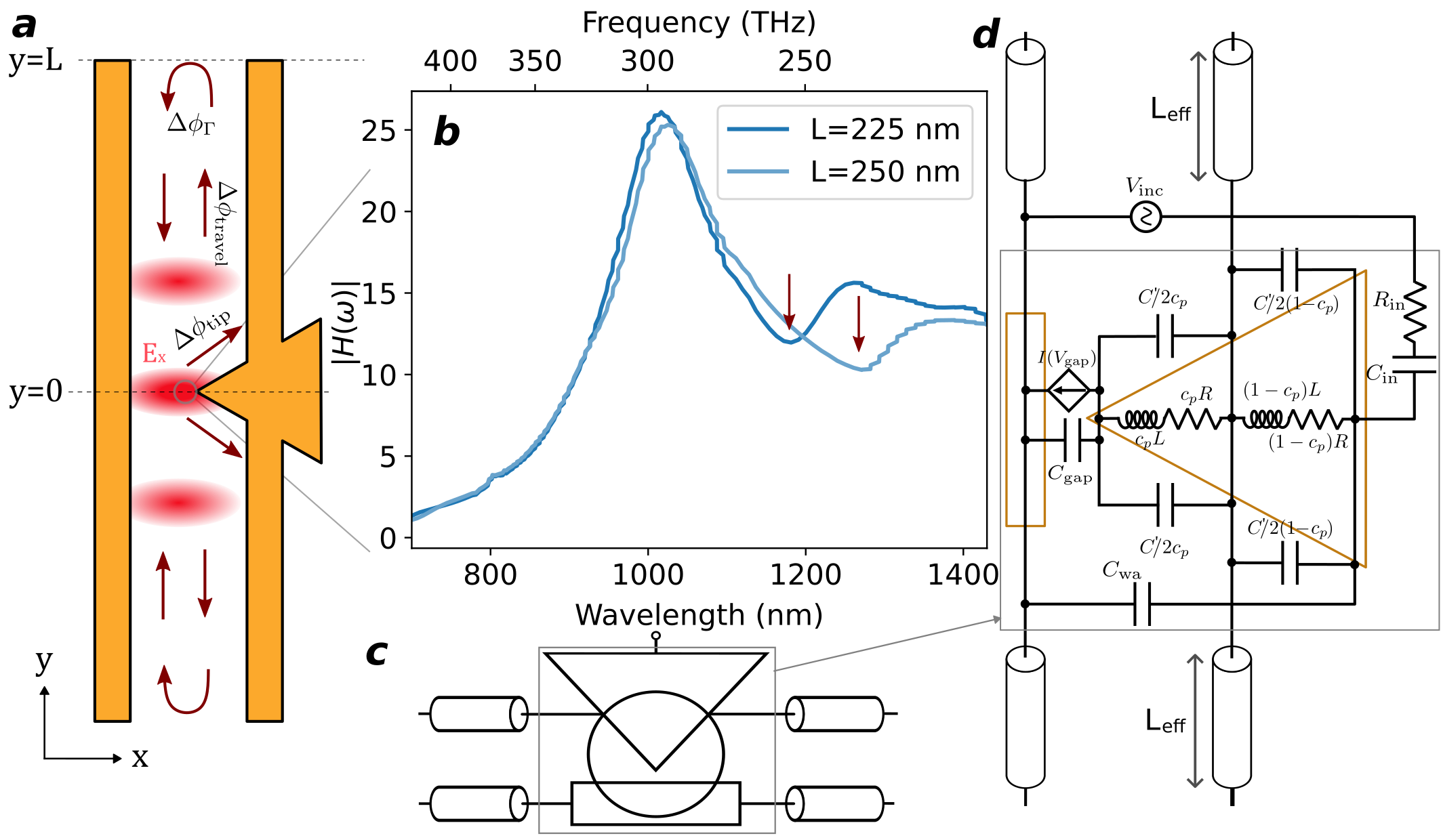}
    \caption{Coupling between an individual antenna structure and nanowires described by modified circuit model and transmission lines. \textbf{a} Layout of a coupled antenna-wire system, which we simulated using an FDTD full-wave solver (MEEP) with varying wire lengths. The signal is injected into the wire cavity by the antenna and experiences a phase shift $\Delta \phi_\text{tip}$ when traveling on the surface of the antenna, in addition to the phase shift $\Delta \phi_\text{travel}$ when traveling on the surface of the straight wires and a phase shift $\Delta \phi_\Gamma$ due to reflection from the termination impedance. \textbf{b} Transfer functions of triangular antennas coupled to wire pairs of different lengths, calculated from FDTD simulations. The transfer function was calculated from the antenna tips. The dips in the transfer function are due to interference from reflected traveling modes. The wavelength locations of the dips depend on the wire length. The frequency of the resonance dip indicates wire properties such as phase velocity and termination load. \textbf{c} Circuit symbol used as a shorthand for the antenna-collector unit cell circuit. \textbf{d} Updated circuit model of antenna and collector accounting for the splitting of body inductance, resistance, and capacitance due to wire connection and interconnects shown as transmission lines. The split is non-uniform and is controlled by a coupling parameter, $c_p \in (0,1)$.}
    \label{fig:wire-char-fdtd-sim}
\end{figure}
To characterize the interaction between the antennas and the nanowires, we produced a set of full-wave electromagnetic simulations of triangular antennas attached to wire pairs of varying lengths. The illustration in \cref{fig:wire-char-fdtd-sim}.\textbf{a} shows a general layout of the structure simulated along with a depiction modes traveling on the wire and getting reflected by the impedance mismatch at the edge. As shown in \cref{fig:wire-char-fdtd-sim}.\textbf{b}, the wavelength/frequency dependent antenna transfer function $H_{FDTD}(\lambda)$ evaluated at the tip of the wire-attached triangular antenna possesses a resonance dip, the wavelength/frequency of which depends on the wire length. The dip is a result of reflected traveling modes interfering destructively with fields at the antenna tip. The total phase shift incurred by a mode traveling from the tip of the triangular antenna through the load and back is:
\begin{equation} 
    \Delta \phi_{\text{ttl}} = 2 \frac{2 \pi L n}{\lambda_0} + \Delta \phi_\Gamma + \Delta \phi_{\text{tip}} = m \pi, \ m \in \{1, 2, 3, 5, \dots\} \label{eq:delta_phi_ttl}
\end{equation}
where $L$ is the length of the wire from the center of the triangular structure to the edge, $n$ is the effective refractive index of the wire surface, $\lambda_0$ is the vacuum wavelength of the mode, $\Delta \phi_\Gamma$ is the phase shift due to the reflection off of the wire edge, and $\Delta \phi_{\text{tip}}$ is the phase shift due to travel from the tip of the antenna to the wire through the triangular structure's width. The total phase accumulated by the traveling mode sums up to an odd multiple of $\pi$ when destructive interference occurs. 

\Cref{fig:wire-char-fdtd-sim}.\textbf{c} shows a circuit symbol which will be used as a shorthand to refer to the antenna-collector circuit model for the duration of the paper. The compact model equivalent to the structure in \cref{fig:wire-char-fdtd-sim}.\textbf{a} and the internal model schematic of the symbol in \cref{fig:wire-char-fdtd-sim}.\textbf{c} is shown in \cref{fig:wire-char-fdtd-sim}.\textbf{d}. The connected wire splits $R_\text{body}$ and $L_\text{body}$ non-uniformly, with the split described by coupling parameter, $c_p$ between 0 and 1. The splitting of the body capacitance, $C_\text{body}$, affects the distance component of the capacitance primarily. The value of $C'_\text{body}$, the equivalent series capacitance of $C'_\text{body}/c_p$ and $C'_\text{body}/(1-c_p)$, is slightly lower than the value of $C_\text{body}$ in the initial circuit model of the antenna unit cell in \cref{fig:unitcell-fitting}.
\begin{figure}[htbp]
    \centering
    \includegraphics[width=1\linewidth]{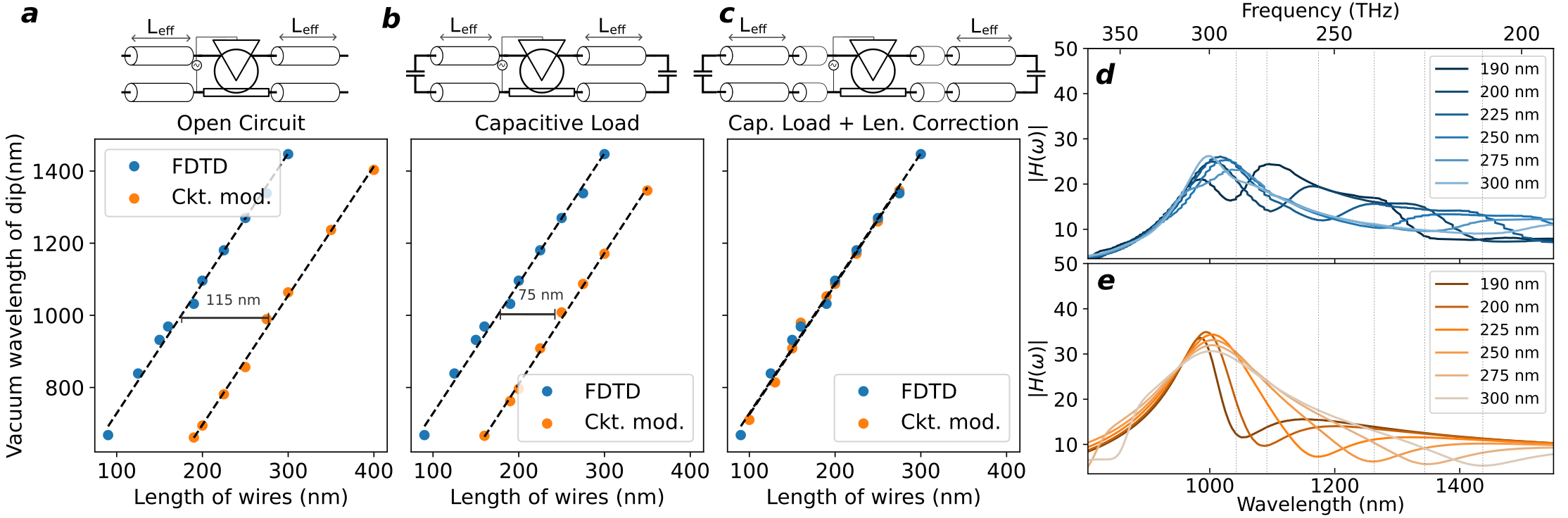}
    \caption{Study of the wavelength location of dips in the transfer function and their rate of change with respect to wire length.  \textbf{a-c} Wavelength of resonance dip in FDTD simulation and compact circuit simulation as a function of wire length for three circuit model cases. \textbf{a} The transmission lines are terminated with an open circuit, leading to a 115 nm mismatch between the FDTD results and the circuit model. Note that both FDTD and circuit model are in agreement over rate of change of resonance dip with respect to wire length, indicating similar phase velocity. \textbf{b} The transmission lines are terminated with a capacitor of 0.468 aF, narrowing the mismatch between the FDTD results and the circuit model to 75 nm. \textbf{c} In addition to the capacitive load of 0.468 aF,  a small transmission line stub of 75 nm is inserted in either end of the antenna, accounting for phase shifts due to antenna coupling effects. \textbf{d} A set of antenna transfer functions calculated from FDTD simulations of triangular nanoantennas coupled to wire-pairs of varying lengths. The transfer functions were calculated at the antenna tips. Dotted vertical lines mark the wavelengths of the interference dips seen in the transfer functions. \textbf{e} A set of transfer functions calculated from circuit solutions of the antenna circuit model coupled to a transmission lines of varying lengths. Dotted vertical lines show agreement between the wavelengths of the dips seen in the circuit model transfer functions and the wavelengths of the dips seen in the FDTD simulated transfer functions.}
    \label{fig:Tline-length-study}
\end{figure}

We implemented the compact model in LTspice with wires modeled as lossy transmission lines with characteristic parameters extracted from FDTD simulations. To overcome time resolution limits of the circuit solver, we introduced a unit transformation described in \cref{sec:spice-transf}. In \cref{fig:Tline-length-study}, we compare the wavelengths of the interference dips from FDTD simulation with the wavelengths of the interference dips from three different sets of compact model circuit simulations implemented in LTspice. The circuit model data in \cref{fig:Tline-length-study}.\textbf{a} was obtained from a set of circuit solution of the antenna model coupled to transmission lines with different effective lengths terminated by an open circuit, as shown in the diagram above \cref{fig:Tline-length-study}.\textbf{a}. The wavelength of the transfer function interference dip is plotted against the wire length for each of the FDTD simulations and the circuit model simulations. A linear trend fit shows a near-linear dependence of the dip wavelength on the length of the wires, indicating that the phase shift in eq.\eqref{eq:delta_phi_ttl} is dominated by the travel along the wire's length. While their slopes match, the FDTD trend line and the circuit model trend line are separated by a constant offset of 115 nm.  

In \cref{fig:Tline-length-study}.\textbf{b}, the circuit model data were obtained from a set of circuit simulations of the antenna model coupled to transmission lines of varying lengths terminated with a capacitive load, the value of which is shown in \cref{fig:transmission-line-char}.\textbf{d}. The wavelength of the interference dip is plotted against the wire length for both the FDTD simulation set and the updated circuit simulation set with linear trend fitting. The slope of the linear trend of the circuit model changes minimally between \cref{fig:Tline-length-study}.\textbf{a} and \cref{fig:Tline-length-study}.\textbf{b}. The separation between the FDTD trend line and the circuit model trend shrinks to a constant offset of 75 nm. The constant offset effect of a capacitive termination on the $\lambda_0$-L relationship is consistent with a simplified small angle approximation of eq.\eqref{eq:delta_phi_ttl} derived in \cref{sec:delta-phi-ttl-deriv}.
 \begin{equation} 
     \lambda_0 = \frac{4\pi(Ln - Z_0 C_L c)}{m \pi - \Delta \phi_{\text{tip}}}, \label{eq:simplified-L-wavelen}
 \end{equation}
where $C_L$ is the capacitance value of the load and $c$ is the speed of light. 

For \cref{fig:Tline-length-study}.\textbf{c}, the circuit model was augmented with a constant length correction term to account for $\Delta \phi_\text{tip}$ in \eqref{eq:delta_phi_ttl}, the wavelength dependent phase shift accumulated by a traveling mode due to traversing the width of the triangular structure away from the tip and onto the wire before propagating down the length of the wire. While the width of the antenna is accounted for in the original transmission line length formulation, the mode travel along the perpendicular component of the triangular structure that is not yet described by the model. The constant length correction section of the transmission line brings the compact model into agreement with the FDTD simulations as seen in \cref{fig:Tline-length-study}.\textbf{c} and in the combination of \cref{fig:Tline-length-study}.\textbf{d} and \cref{fig:Tline-length-study}.\textbf{e}. We emphasize this is a fixed correction that does not depend on the transmission line length. The wavelength dependent transfer functions calculated from FDTD simulation set are plotted in \cref{fig:Tline-length-study}.\textbf{d}. The vacuum wavelengths of the destructive interference dips are marked with vertical dotted lines. The wavelength dependent transfer functions calculated the circuit simulation set of the structure in \cref{fig:Tline-length-study}.\textbf{c} are plotted in \cref{fig:Tline-length-study}.\textbf{e}. The dips are in agreement with those from the FDTD simulations.

\begin{figure}[htbp]
    \centering
    \includegraphics[width=0.95\linewidth]{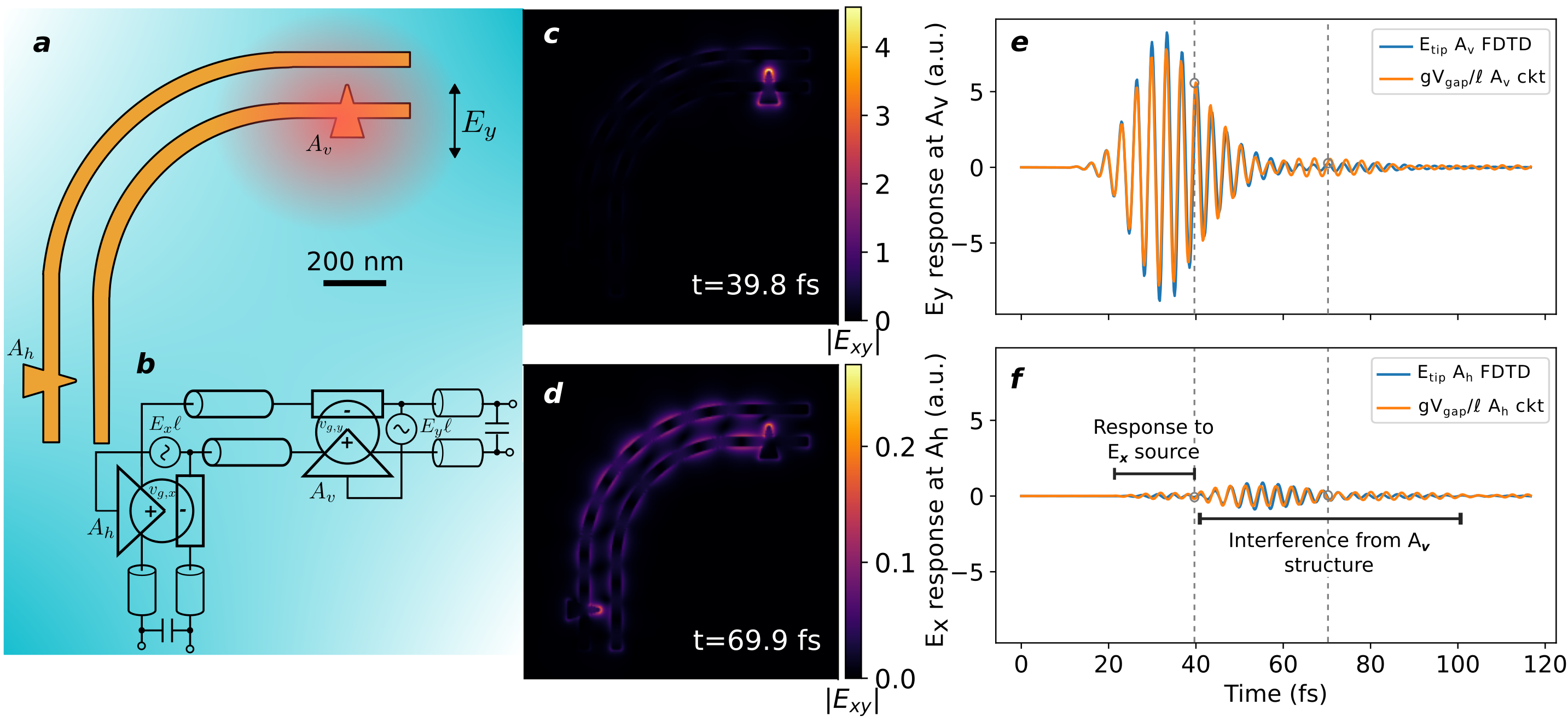}
    \caption{The compact model is effective in predicting the dynamics of a systems with two polarization-sensitive antennas coupled through a two-wire transmission line. \textbf{a} Layout of a system containing polarization-sensitive antennas connected with a two-wire plasmonic transmission line. An FDTD simulation was used to calculate the full structure's electromagnetic response to a local Y-polarized beam incident on the vertical antenna. \textbf{b} Compact circuit model corresponding to the coupled antenna system shown in \textbf{a}. \textbf{c-d} Time snapshots of the electrical field magnitude in an XY plane slice of the FDTD simulation of the structure in \textbf{a}. \textbf{c-d} Snapshots of the electric field profile along the structure at \textbf{c} 39.8 fs and at \textbf{d} 69.9 fs.  \textbf{e-f} Comparison of simulated electric fields at the tips of \textbf{e} the active vertical antenna and \textbf{f} the passive horizontal antenna.  The curves shown are from full wave FDTD simulation (blue curve) and compact model solution (orange curve). Compact model fields were again only scaled by the geometrical enhancement factor of 7.22 as derived from the unit cell model.  In \textbf{f} the times corresponding to the field snapshots shown in \textbf{c} and \textbf{d} are indicated by vertical dashed lines.}  
    \label{fig:coupled-xy-antennas}
\end{figure}

Next, we performed an FDTD simulation of a system of two triangular nanoantennas oriented at 90 degrees from one another and connected by a pair of rounded wires, as shown in \cref{fig:coupled-xy-antennas}.\textbf{a}, to examine the circuit model's efficacy in predicting the coupling between optically electrically-connected excited nanostructures. The vertical antenna in \cref{fig:coupled-xy-antennas}.\textbf{a} is labeled $A_v$, and the horizontal antenna is labeled $A_h$. The incident optical input was polarized in the Y direction, along the axis of $A_y$, and had a localized rectangular beam profile of size $300 \text{ nm} \times 500 \text{ nm}$ centered on $A_v$. While the input was defined as local and Y-polarized, an FDTD simulation of an empty simulation cell with an identical source beam revealed a small, yet detectable, X-polarized electric field component incident in the location of $A_h$ due to the radiation pattern of the source dipole.

The circuit model equivalent is shown in \cref{fig:coupled-xy-antennas}.\textbf{b}. Transmission line stubs with capacitive termination were added at to match the stubs present in the FDTD simulation. Polarization sensitivity is accounted for in the circuit model with the independent voltage sources connected to individual antennas. The amplitude of the signals in the voltage sources attached to $A_v$ and $A_h$ matched the relative amplitudes of the incident Y polarized pulse and the X polarized background radiation component from the FDTD simulation.

\Cref{fig:coupled-xy-antennas}.\textbf{c-d} show time snapshots of the amplitude of the planar XY electric fields calculated by the FDTD simulation. \Cref{fig:coupled-xy-antennas}.\textbf{c} shows a snapshot at time $39.8$ fs when the fields are strongest at the tip of $A_v$ as the resonance response of the Y-polarized incident pulse had barely started to decay. At time $69.9$ fs, the resonant response of $A_v$ had decayed. As shown in \cref{fig:coupled-xy-antennas}.\textbf{d}, standing waves have formed on the rounded wires as a result of interferences of reflected traveling modes. The traveling modes mediate optical coupling between $A_v$ and $A_h$. 
 
 \Cref{fig:coupled-xy-antennas}.\textbf{e} compares $E_{y,\text{tip}}$, the FDTD calculated Y component of the electric field evaluated at the tip of $A_v$ (in blue) with $g V_\text{gap}/\ell$, the circuit model solution for the geometrically enhanced and length scaled voltage across the gap of $A_v$ (in orange). The circuit and FDTD models agree very well aside from the peak field obtained from the circuit model being slightly lower than the peak field calculated by the FDTD simulation. \Cref{fig:coupled-xy-antennas}.\textbf{f} compares $E_{x,\text{tip}}$, the X component of the FDTD simulated electric field evaluated at the tip of $A_h$ (in blue) with $g V_\text{gap}/\ell$, the circuit model solution for the geometrically enhanced and length scaled voltage across the gap of $A_h$ (in orange). The structure of both the FDTD result and the circuit model result consists of two pulses; a low amplitude one concurrent with the resonance response of $A_v$, and a higher amplitude one delayed by 20 fs from the $A_v$ resonance peak. The earlier pulse is driven by the weak X component of the radiation from the dipole generating the local Y-polarized source. The later pulse is caused by interference from the $A_v$ resonance response which traveled through the nano-wires to excite $A_h$. The 20 fs delay between the peaks of the $A_v$ resonance and the peak of the interference signal at $A_h$ is close to the 16 fs expected time of travel of a surface mode with phase velocity $v_p = 0.35c$ propagating on a transmission line with effective length of 1663 nm. 

\subsection*{Phase II: Analysis of Charge Transfer Characteristics and Simulated Cross-Correlation Measurement for Benchmarking} 
In the presence of electrical fields that obey the Keldysh requirement of $\gamma_k <1$ for a given metal, charge emission from the metal is dominated by a field-driven tunneling process \cite{hommelhoff_field_2006}. Electron emission from triangular antennas is rectified, occurring during the negative half-cycles of the driving fields (defined with respect to the outward pointing normal to the surface) \cite{putnam_optical-field-controlled_2017} and emission is concentrated at the tip, where the geometric field enhancement is strongest \cite{Buckley:21}. Experimental work has demonstrated that charge emission from triangular nanostructures follows a mixed mode trend with a thermal Schottky process dominating at low field strengths and  a Fowler-Nordheim process dominating at higher field strengths \cite{turchetti_electron_2022}. In the circuit model, electron emission is modeled using a rectified voltage dependent current source stretching from the emitter tip node to the collector node, parallel to the gap capacitor, as shown in \cref{fig:Tline-length-study}.\textbf{c}. The tunneling current rate is set by a weighted sum, such that Schottky emission dominates in low-field strengths and Fowler-Nordheim emission dominates at high fields. The field dependent emission current is
\begin{equation}
    \Gamma(E_\text{tip}) = \Gamma(g V_\text{gap}/\ell) = w_{fn}A_\text{eff} J_{fn}\left(\frac{g \min(V_\text{gap}, 0)}{\ell}\right) + w_{s}A_\text{eff} J_{s}\left(\frac{g \min(V_\text{gap},0)}{\ell}\right),
\end{equation}
where $w_{fn}$ and $w_{s}$ are the weighted sum parameters, $A_\text{eff}$ is the effective area of the emission region, $J_{fn}$ and $J_s$ are the Fowler-Nordheim and the Schottky field dependent electron emission current density rates functions respectively. The gap voltage value proportional to the electric field at the antenna tip such that $gV_\text{gap} /\ell$. Following the formulation in ref \cite{turchetti_electron_2022}, the Fowler-Nordheim emission rate is defined as
\begin{equation}
   J_{fn}(g V_\text{gap}/\ell) = \phi^{-1} \left( g V_\text{gap} \right/\ell)^2 \exp \left( -b \frac{ \phi^{3/2} \ell}{g V_\text{gap}} v(f) \right),
\end{equation}
where $\phi$ is the work function and $b = 6.8309 \text{eV}^{-3/2}\text{V}/\text{nm}$. The exponent includes a correction term $v(f)=1-f +f\ln(f)/6$ with $f=g V e^3/(4 \pi \epsilon d)\phi^2$  where $e$ is the elementary charge and $\epsilon$ is the permittivity of the metal. The Schottky emission rate is
\begin{equation}
    J_{s} (g V_\text{gap}/\ell) = T^2 \exp \left(\frac{-\phi +e/2}{k_B T} \sqrt{\frac{e g V_\text{gap}}{ \pi \epsilon \ell}} \right)
\end{equation}
where $k_B$ is the Boltzmann constant and $T$ is the temperature of the metal. 

The tunnel-emitted electrons enter the gap with zero kinetic energy, but quickly gain momentum due to the ponderomotive force induced by the electric field spatial gradient \cite{Lozovski2022, RevModPhys.72.545} as well as externally applied bias. In structures with gaps that are 10 nm or smaller, the emitted electrons get swept across the gap in sub-cycle times \cite{ludwig_sub-femtosecond_2020}. In devices with larger gaps, a time delay of tens of fs might need to be included in the model to account for the transport time of the electrons across the gap. Setting the exact delay time will require calibration by experiment. 

Another consideration is whether the model changes due to the electron plasma generated in the nanometer-scale gap from emission.  The optical properties of the moving electrons in the structure's gap can be described with a kinetic inductor placed in series with the emission current source. The inductor is derived based on the application of the Drude model to the electron plasma in the gap in \cref{sec:append-optical-effects-of-plasma}. Experimental work by Bionta et al. used a pump-probe measurement to sample sub-femtosecond dynamics from a triangular antenna. They showed that with a large pump pulse and a low-amplitude variable-delay probe pulse, the time-averaged current from a triangular antenna followed the scaled temporal profile of $E(\tau)\ast h_\text{antenna}$, where $E(t)$ was the probe pulse, $\tau$ was the delay, and $h_\text{antenna}(t)$ was the impulse response of the antenna, the time domain analogue to the transfer function calculated by an FDTD simulation \cite{bionta_-chip_2021}. That result indicates the electron plasma in the gap does not have a significant back-action affecting the resonance of the antenna or the emission from it. Optical simulations of the plasma in the gap described in \cref{sec:append-optical-effects-of-plasma} confirm the same. Thus, we may exclude the kinetic inductance of the vacuum gap from the circuit model.

%% file: Application_example.tex
\begin{figure}[htbp]
    \centering
    \includegraphics[width=0.96\linewidth]{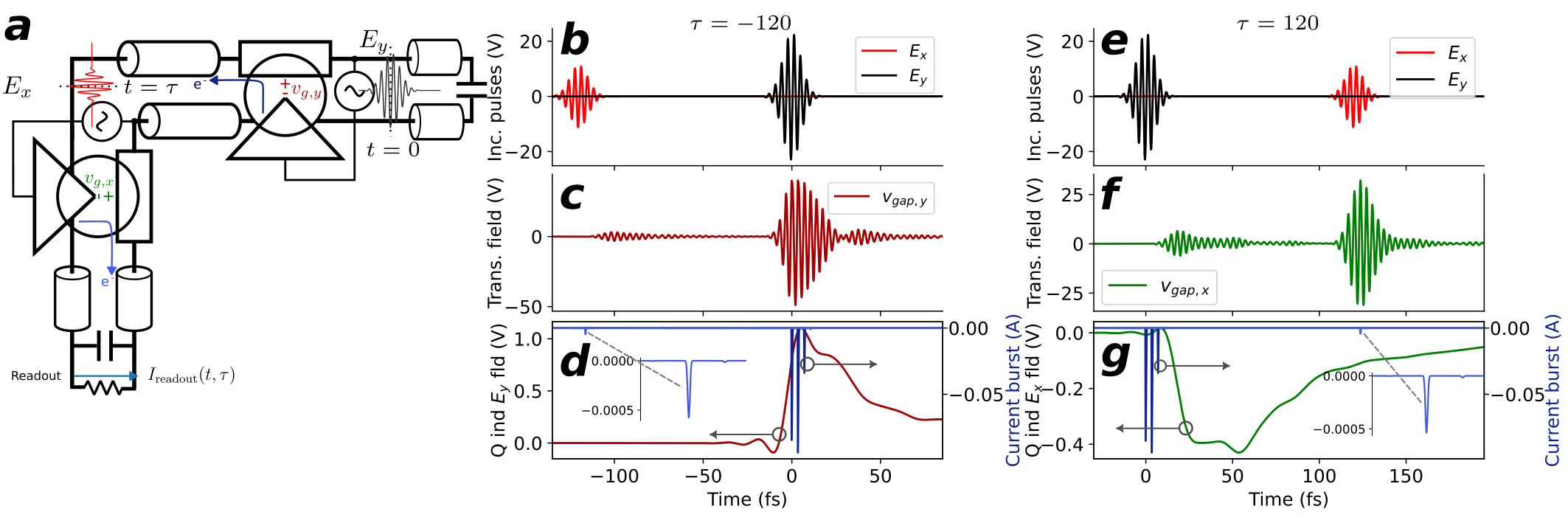}
    \caption{Transient fields calculated by a circuit simulation of a polarization-sensitive coupled antenna system. When $\tau < 0$ the low amplitude $E_x$ probe pulse arrives before the high amplitude $E_y$ pump. When $\tau > 0$ the high amplitude $E_y$ pump pulse arrives before the low amplitude $E_x$ probe. \textbf{a} Compact model representation of the layout. The voltage source representing the Y (X) polarized pulse acting on the vertical (horizontal)  antenna has a temporal gaussian envelope pulse profile centered at time $t=0$ ($t= \tau$). The readout resistor models the electronic readout with its series resistance. \textbf{b} Transient time domain fields of input pulses with the pump pulse centered at $t = 0$ fs, with the negative delay of $120$ fs, te probe is centered at $t=-120$ fs. \textbf{c} The transient time dependent field induced at the tip of $A_v$. The oscillations in the field between -110 and -60 fs are due to interference from $E_x(t)\ast h_\text{antenna}(t)\ast h_\text{Tline}(t)$, the resonance response of $A_h$ to $E_x$. \textbf{d} (left y-axis) Fields induced at $A_v$ due to slow moving charges obtained from applying a low pass filter to $V_{\text{gap},y}$, the gap voltage at $A_v$. (right y-axis) Electron emission current bursts. \textbf{e} Transient time domain fields of input pulses with the pump pulse centered at $t_0=0$. With a positive delay $\tau = 120$ fs, the probe centered at $t=120$ fs. \textbf{f} The transient time dependent field induced at the tip of $A_h$. The oscillations in the field between 0 and 70 fs are due to interference from the resonance response of $A_v$ to $E_y$. \textbf{g} (left y-axis) Fields induced at $A_h$ due to slow moving charges obtained from applying a low pass filter to $V_{\text{gap},x}$, the gap voltage at $A_h$. (right y-axis) Electron emission current bursts.}
    \label{fig:benchmark-transient}
\end{figure}

With the charge emission source now in place, we then turned our efforts towards benchmarking the model.  Unfortunately, there are no former experiments or established models to turn to in order to provide a trusted benchmark. Given this lack of an established experimental reference, we rather used our model to propose a cross correlation measurement that could be used to quantitatively verify our model through future experiment.  We choose a cross-correlation measurement scheme as it enables time-resolved probing of the femtosecond dynamics with  much lower bandwidth readout electronics given that direct measurement of the femtosecond dynamics at play is not available. 

For this proposed cross-correlation scheme, we focus on the two-antenna system whose electromagnetic properties were examined in \cref{fig:coupled-xy-antennas}.\Cref{fig:benchmark-transient}.\textbf{a}, shows a schematic of the circuit model representation of the two-antenna system. The readout resistor represents electronic readout with known impedance. A short Y-polarized pulse, $E_y(t)$, centered at time $t=0$ with central wavelength $1.062 \mu$m, a $5.1$ fs full-width-half-maximum (FWHM), and peak field of 0.36 GV/nm, is represented by a voltage source attached to $A_v$.  The voltage source attached to $A_h$ represents a X-polarized pulse, $E_x(t, \tau)$, with the same central wavelength and duration as $E_y(t)$, but centered at a variable time $t=\tau$ and with a peak field of 0.17 GV/nm. 

\Cref{fig:benchmark-transient}.\textbf{b-d} shows transient fields and electron emission current bursts for a delay of $\tau = -120$ fs. \Cref{fig:benchmark-transient}.\textbf{b} shows the waveforms of the $E_x(t, \tau)$ probe pulse and of the $E_y(t)$ pump, which arrives 120 fs later. The probe pulse, $E_x (t, \tau)$, induces a resonance in $A_h$, with fields at the tip  $E_{\text{tip},x} = E_x (t, \tau) \ast h_\text{antenna} \propto (t) V_{\text{gap}, x}$, where $h_\text{antenna} (t)$ is the impulse response of the antenna. Interference modes propagate through the transmission lines as a result. 

\Cref{fig:benchmark-transient}.\textbf{c} shows $V_{\text{gap},y} \propto E_{\text{tip}, y}$,  the voltage across the gap of $A_v$. Before $E_y (t)$ strikes at $t=0$, the fields at the tip of $A_v$ are entirely a result of the interference from $A_h$, such that $E_{\text{tip}, y} = E_x(t, \tau) \ast h_\text{antenna}(t) \ast h_\text{Tline}$, where $h_\text{Tline}$ is the impulse response of the transmission line. Upon the arrival of $E_y (t)$, $E_{\text{tip}, y}$ becomes a superposition of the resonance of $A_v$ and the interference from $A_h$, $E_{\text{tip}, y} = E_x(t, \tau) \ast h_\text{antenna}(t) \ast h_\text{Tline} + E_y(t)\ast h_\text{antenna}$. As the delay $\tau$ is varied, $E_x(t, \tau) \ast h_\text{antenna}(t) \ast h_\text{Tline}$ modulates the tunneling current from $A_v$. 

Charge emission from a structure generates slowly-varying field components which add a bias to the fields at the tip of the antennas. The left y-axis of \cref{fig:benchmark-transient}.\textbf{d} shows the charge-related slow bias component isolated from $V_{\text{gap}, y}$ using a low-pass filter. The right y-axis of \cref{fig:benchmark-transient}.\textbf{d} shows the instantaneous tunneling currents from both $A_v$ and $A_h$. 

\Cref{fig:benchmark-transient}.\textbf{e-g} shows transient fields and electron emission current bursts for a delay of $\tau = 120$. \Cref{fig:benchmark-transient}.\textbf{e} shows the waveforms of the Y-polarized pump, $E_y (t)$, and the delayed X-polarized probe, $E_x (t, \tau)$. The pump, $E_y(t)$, induces a resonance response $E_{\text{tip}, y} = E_y(t) \ast h_\text{antenna}$ at the tip of $A_v$ with interference modes that travel to $A_h$. \Cref{fig:benchmark-transient}.\textbf{f} shows $V_{\text{gap}, x}$, the voltage across the gap of $A_h$, which has two components. The pulse between 0 to 100 is $ E_y(t) \ast h_\text{antenna} (t) \ast h_\text{Tline}(t)$, the interference from $A_v$. The pulse at $t>100$ is $E_x(t, \tau) \ast h_\text{antenna} (t)$, the resonance response of $A_h$ to the $E_x(t, \tau)$ pulse. The charge emitted from $A_v$ while it is driven by $E_y(t)$ propagates through the system and applies an additional slow bias to $A_h$. The left y-axis of \cref{fig:benchmark-transient}.\textbf{g} shows the charge-related low frequency component of $V_{\text{gap}, x}$. The right y-axis shows the instantaneous tunneling currents from $A_v$ and $A_h$. 

\begin{figure}[htbp]
    \centering
    \includegraphics[width=0.96\linewidth]{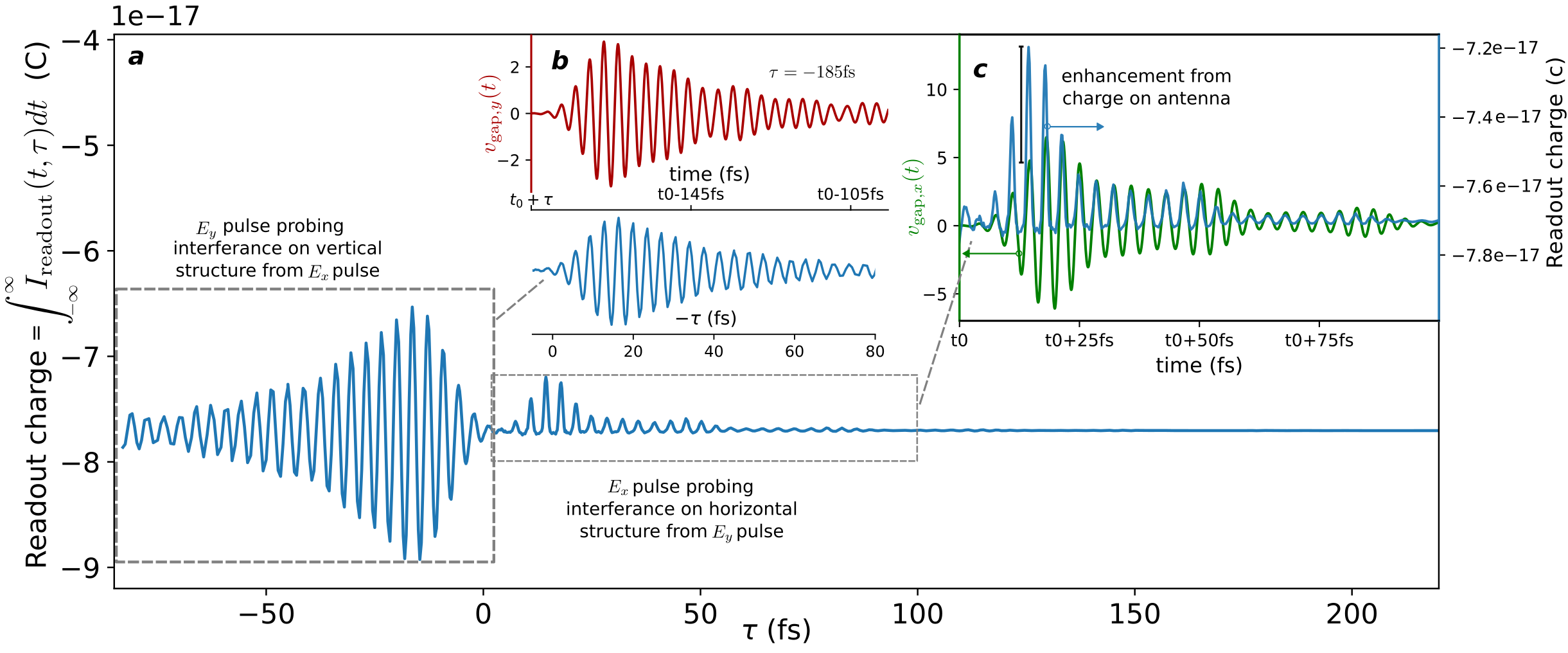}
    \caption{Results of a simulated cross correlation measurement proposed as a benchmarking experiment for the non-linear charge dynamics in the compact model. \textbf{a} The simulated total readout charge $Q_\text{read} = \int \Gamma (E_{\text{tip}, y}(t))dt - \int \Gamma (E_{\text{tip}, x}(t))dt$, defined as the time integral of the instantaneous current through the readout resistor, as a function of $\tau$, the delay between the center envelopes of the $E_y(t)$ and $E_x(t, \tau)$ pulses. \textbf{b} Comparison of the instantaneous voltage across the gap of $A_v$ in a simulation instance of $\tau = -185$ fs with readout charge for $-85 <\tau <0$ fs. The shape of the read-out charge vs $-\tau$ in that interval matches $E_x(t, \tau) \ast h_\text{antenna} \ast h_\text{Tline}$, the interference fields induced at $A_v$ due to coupling from $A_h$. The comparison shows that the cross correlation can capture femtosecond electromagnetic and charge transfer dynamics with conventional electronics. \textbf{c} A comparison of instantaneous voltage across the gap of $A_h$ for $\tau = 180$ fs (left y-axis) with readout charge for $0<\tau < 100$ fs (right y-axis). The enhancement at $\tau=12$ fs is caused by the slowly varying bias induced on $A_h$ by the charge emitted from $A_v$. }
    \label{fig:benchmark-charge-coupling}
\end{figure} 

 We varied $\tau$ from -83 to 220 fs, with 625 sub-femtosecond non-uniformly spaced delay time steps. For each value of $\tau$, we ran the transient circuit simulation until the charge had settled. In post processing we integrated the current through the readout to obtain the total readout charge. \Cref{fig:benchmark-charge-coupling}.\textbf{a} shows the readout charge as a function of the delay of $E_x(t, \tau)$. Since the currents generated by $A_v$ and $A_h$ enter the readout from the opposite directions, the total readout charge is $Q_\text{read} = \int \Gamma (E_{\text{tip}, y}(t))dt - \int \Gamma (E_{\text{tip}, x}(t))dt$, the total charge emitted from $A_v$ minus the total charge emitted from $A_h$. \Cref{fig:benchmark-charge-coupling}.\textbf{b} zooms in on the $\tau <0$ section of the readout charge and compares it to the instantaneous $V_{\text{gap},y}$ for a case of $\tau = -185$ fs. When $\tau <0$, the emission from $A_v$ is modulated by the interference from $A_h$ and the total charge can be expressed as 
 \begin{equation} \label{eq:ch6-current-tau-neg}
       Q_\text{read} (\tau) =  \int \left( \Gamma(E_y (t)\ast h_\text{antenna} (t) + E_x(t, \tau) \ast h_\text{antenna} (t)\ast h_\text{Tline}(t))
    -   \Gamma (E_{\text{tip}, x}(t)) \right) dt.
\end{equation}
 Following the small signal derivation from Bionta et al. \cite{bionta_-chip_2021}, we get
\begin{equation}
Q_\text{read}(\tau) = \int \left( \Gamma(E_y(t)\ast h_\text{antenna}(t))  + \frac{d \Gamma}{dE}\bigg\vert_{E_y(t)\ast h_\text{ant.} } E_x(t,\tau) \ast h_\text{antenna}(t)\ast h_\text{Tline} (t)\right) dt - Q_h,
\end{equation}
where $Q_h$ is the negligible charge emitted from $A_h$ when $\tau < 0$.  The integral is then resolved into two components: (1) a delay-independent offset $\int \Gamma(E_y(t)\ast h_\text{antenna}(t))dt$ equal to the charge expected from $A_v$ in the absence of a probe pulse and (2) a component proportional to $E_x(-\tau) \ast h_\text{antenna}\ast h_\text{Tline}$. The delay-dependent component of the readout charge probed the sub-femtosecond dynamics of the interference from $A_h$ seen at $A_v$.

\Cref{fig:benchmark-charge-coupling}.\textbf{c} compares the $\tau > 0$ section of the readout charge to the instantaneous field at the tip of $A_h$ when $\tau = 120$. The blue trace (referenced to the right y-axis) shows a zoomed in section from $Q_\text{read}(\tau)$ and the green trace shows $V_{\text{gap},x} \propto E_{\text{tip}, x}$ (left y-axis). 
\begin{equation}
    \begin{split}
        Q_\text{read}(\tau) \propto  &\int \Gamma(E_y (t)\ast h_\text{antenna}(t))dt \\
    - &\int \Gamma(E_x(t-\tau)\ast h_\text{antenna} (t)+ E_y(t)\ast h_\text{antenna} (t)\ast h_\text{Tline}(t) + E_{Q, A_v}(t))dt.
\end{split}
\end{equation}

In addition to the optical interference, $A_v$ produces significant charge which results in an additional slowly varying bias, $E_{Q, A_v} (t)$. The magnitude of the $E_x (t, \tau)$ pulse was set such that $E_x (t, \tau) \ast h_\text{antenna}$ cannot drive significant charge emission on its own. However, the superposition with $E_y (t)\ast h_\text{antenna} \ast h_\text{Tline} + E_{Q, A_v}(t)$ is sufficient to generate noticeable charge emission from $A_h$.  When $E_x(t, \tau) \ast h_\text{antenna}$ and $E_y (t)\ast h_\text{antenna} \ast h_\text{Tline}$ add constructively, the emission from $A_h$ is enhanced. Otherwise, the emission from $A_h$ is suppressed and $Q_\text{read}$ settles to $\int \Gamma (E_y (t)\ast h_\text{antenna})dt$. The enhancement due to $E_{Q, A_v}(t)$ is marked in \cref{fig:benchmark-charge-coupling}.\textbf{c}. After the charge dies out, the readout charge tracks the upper envelope of $E_y(t)\ast h_\text{antenna} (t)\ast h_\text{Tline}(t)$.
The lifetime of the charge in the gap can be extended by increasing the readout resistance, thus increasing the RC time constant of the gap and the surrounding regions.

The circuit simulation of the system reveals that cross correlation measurements can produce a unique fingerprint of the system and to probe transient fields with femtosecond resolution even with slow readout electronics. Such cross-correlation fingerprints from future experimental measurements can then be used as a benchmark to finalize and quantitatively verify the circuit models.  

%% file: conclusions.tex
This work introduces a framework for describing nanoplasmonic structures and their interconnects using compact circuit models which go beyond previous efforts in the field with their ability to (1) predict electric field values at specific locations on individual structures, (2) describe coupling between structures through connecting wires, and (3) account for non linear charge emission as well as charge transfer and charge-mediated coupling. The electromagnetic component of the circuit model was benchmarked against full-wave FDTD simulations for individual antennas, nanowire interconnects, and systems of two antennas coupled through rounded nanowire interconnects. 

The connecting wires in \cref{fig:coupled-xy-antennas,fig:benchmark-charge-coupling} were rounded to allow the propagating modes to travel with low loss. That layout might not be practical for a system with other fabrication constraints, which might need to make use of a tighter layout and sharper corners. In follow-up work, the model can be augmented to account for the radiation loss of sharp corners and near field coupling of tightly arranged elements. 

To provide a future benchmark for the complete model, we proposed a cross-correlation measurement that maps the femtosecond-scale electromagnetic and charge dynamics to low-frequency, time-averaged charge output that can be measured with conventional electronics. Using our model, we show how such cross-correlation measurements contain unique fingerprints of these dynamics and will provide the needed information to quantitatively assess the model's ability to capture electromagnetic and non-linear charge-induced coupling. 

A direct advantage of the compact model is the speed-up it provides for electromagnetic simulations. The electromagnetic FDTD simulation of the coupled antenna system studied in \cref{fig:coupled-xy-antennas} required 10 hours to run on a single thread on a standard desktop PC. In comparison, the circuit model simulation for the same system required only 36 seconds on a comparable computer. However, advantages extend beyond the speed-up of electromagnetic simulations. 

First, the models provide a flexible ecosystem for device design using tools already familiar to the electronics community. The circuit model enables scalable simulation of systems of nanostructures with non-linear charge emission and with charge-mediated coupling, which is not yet available in a single scalable package. Scalable simulations significantly improve the efficiency of device design that relies on iterative simulation loops of incrementally changing structures. 

Second, the model can be used to aid in understanding of the underlying device physics. One can use the models to study systems and to predict their responses to incident pulses with different amplitudes, carrier-envelope phase, or different repetition rates. Beyond simply predicting experimental results, the model facilitates understanding of the phenomena driving experimental results. The cross-correlation study shown in \cref{fig:benchmark-charge-coupling} produced a trace of time averaged current as a function of probe-pulse delay which is experimentally observable, allowing direct comparison of simulation and experiment. The time-averaged readout of cross correlation experiments captures femtosecond dynamics which are too fast to observe directly with low bandwidth readouts. The circuit simulation provides direct access to variables with femtosecond dynamics such as the instantaneous fields on antenna tips. Using the simulation, we directly compared the instantaneous variables in a transient circuit simulation to the dynamics captured in the time-averaged readout from the full simulation set. We were able to confirm that the time-averaged cross correlation measurement readout of the two antenna system captures the optical interference of an antenna resonance which propagated through the system to modulate the emission of the other antenna. The simulation set additionally indicated that charge emitted from one antenna can get temporarily trapped on another electrically-connected antenna, modulating its emission. 

Followup work can build on the charge probing demonstration to develop a PHz memory device in which charge trapped on a structure encodes a state. In that device, the write operation would be performed by the pulse that generates the charge and the read operation would be performed by a small signal pulse that probes the charge and delivers it to the readout \cite{lee_model_2018, bechhofer_circuit_2023}. This framework can be extended to a multi-stage memory devices in which a state, encoded by charge, is serially shifted from one stage to the next by a sequence of ultrashort pulses in a scheme that resembles a shift register. A framework like that could enable a PHz memory and computational architecture with data that is serially shifted to populate a buffer array which would be measured in its entirety in parallel by low bandwidth electronics, enabling PHz on-chip processing with amortized PHz-scale electronic readout.  

%% file: appendix.tex
\subsection{Analytical Circuit Transfer Function Derivation} \label{sec:analytical-transfxn-deriv}
The impedance of the antenna body is given by a parallel combination of the body capacitor with the series combination of body inductance and body resistance. $Z_{\text{body}} = (R_{\text{body}} + L_{\text{body}}) \parallelsum \frac{-j}{\omega C_{\text{body}}}$. The gap impedance is solely defined by the gap capacitance, $Z_{\text{gap}} = \frac{-j}{\omega C_\text{gap}}$. Likewise, for the impedance of the wire to antenna-base capacitance  $Z_{\text{wa}} = \frac{-j}{\omega C_\text{wa}}$. The combined impedance of the antenna structure is $Z_{\text{struct}} = (Z_{\text{body}} + Z_{\text{gap}}) \parallelsum Z_{\text{wa}}$. The input impedance of the electromagnetic excitation is $Z_{\text{in}} = R_\text{in} + \frac{-j}{\omega C_{\text{in}}}$. 

The transfer function is defined as $H(\omega) = \frac{V_{\text{gap}}}{V_\text{opt}}$. We use the combined effective impedance to get an analytical expression for $H(\omega)$ from the circuit parameters. 
\begin{equation}
    I_{\text{opt}} = \frac{V_{\text{opt}}}{Z_\text{struct} + Z_\text{in}}
\end{equation}
\begin{equation}
    I_{\text{gap}} = \frac{I_{\text{opt}}Z_\text{struct}}{Z_\text{body}+Z_\text{gap}}
\end{equation}
\begin{equation}
      V_{\text{gap}} = I_{\text{gap}}  Z_{\text{gap}} = \frac{Z_{\text{gap}} Z_{\text{struct}}}{Z_{\text{body}}+Z_{\text{gap}}} I_{\text{opt}} = \frac{Z_{\text{gap}} Z_{\text{struct}}}{Z_{\text{body}}+Z_{\text{gap}}} \frac{V_{\text{opt}}}{Z_\text{struct} + Z_\text{in}}
\end{equation}
The transfer function can now be expressed as 
\begin{equation}
    H(\omega) = \frac{Z_{\text{gap}} Z_{\text{struct}}}{Z_{\text{body}}+Z_{\text{gap}}} \frac{1}{Z_\text{struct} + Z_\text{in}}. 
\end{equation}
Substituting in the impedances from above yields:
\begin{equation}
\begin{split}
H(\omega) = &\frac{\frac{-j}{\omega C_\text{gap}}}{(R_{\text{body}}+j\omega L_\text{body}) \parallelsum \frac{-j}{\omega C_{\text{body}}} + \frac{-j}{\omega C_{\text{gap}}}} \\
&\frac{\frac{-j}{\omega C_{\text{wa}}}\parallelsum \left((R_{\text{body}}+j\omega L_\text{body}) \parallelsum \frac{-j}{\omega C_{\text{body}}} +\frac{-j}{\omega C_\text{gap}}\right)}{\frac{-j}{\omega C_{\text{wa}}}\parallelsum \left((R_{\text{body}}+j\omega L_\text{body}) \parallelsum \frac{-j}{\omega C_{\text{body}}} +\frac{-j}{\omega C_\text{gap}}\right) + R_{\text{in}} + \frac{-j}{\omega C_{\text{in}}}}.
\end{split}
\end{equation}
The final expression is a complex function of frequency and circuit model parameters. 

\subsection{Unit Cell Simulations} \label{sec:fdtd-sims-unit-cell}
\begin{figure}[htbp]
    \centering
    \includegraphics[width=0.85\linewidth]{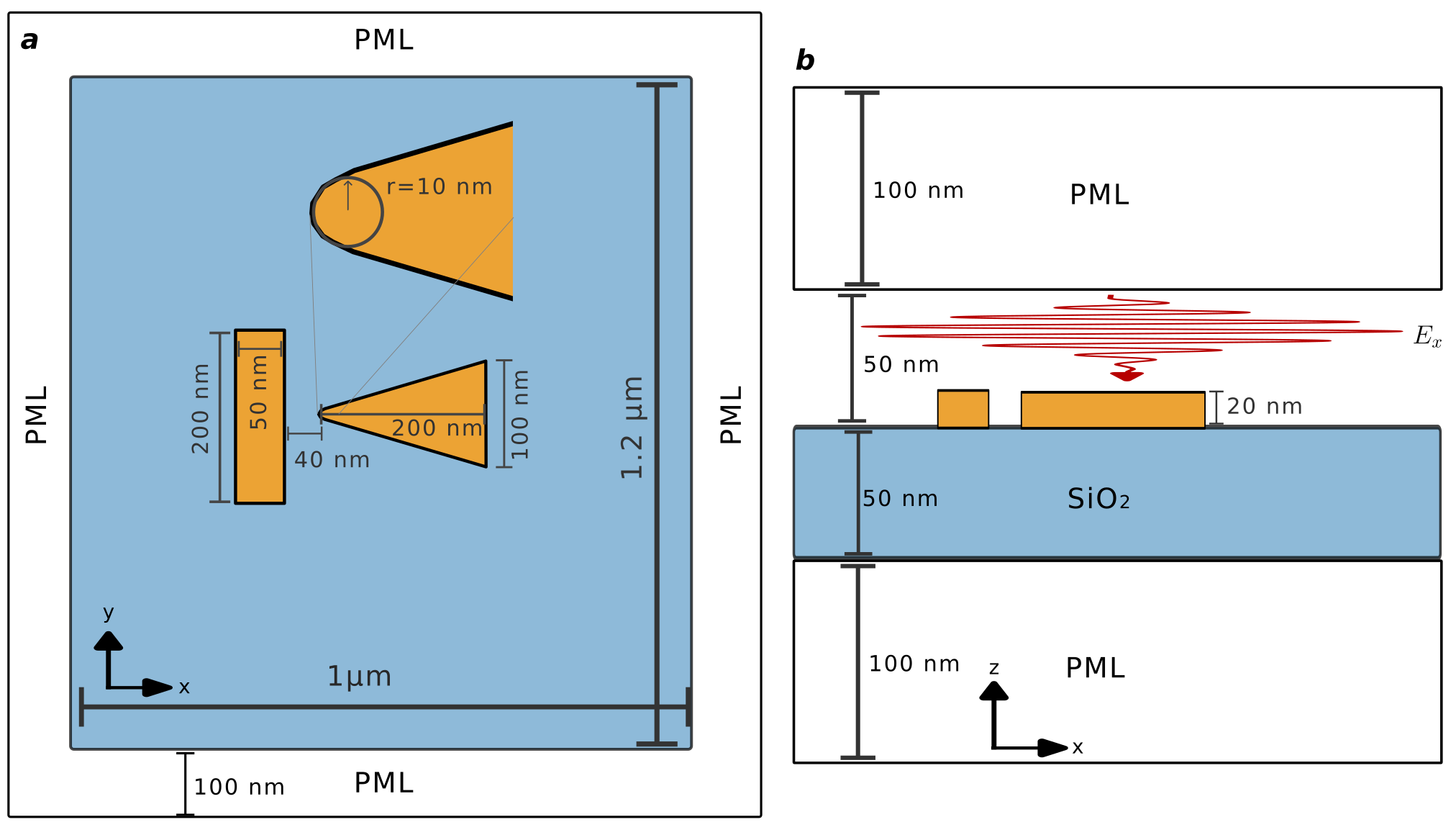}
    \caption{Layout of unit cell simulation in FDTD simulator. \textbf{a} Top down view of a $1 \mu \text{m} \times 1.2 \mu \text{m}$ cell surrounded by 100 nm of PMLs from all sides. The gold structure consists of a 200 nm long triangle and a 50 nm wide wire with a 40 nm gap in between. \textbf{b} Cross section in the xz plane of the simulation cell. The gold structure is placed on a layer of glass with the incident field coming from the top of the simulation cell. The source is placed right below the PML layer.}
    \label{fig:simulation-layout}
\end{figure}
The FDTD simulations of the antenna-wire pairs were executed in a three dimensional simulation cells of size $1 \mu \text{m} \times 1.2 \mu \text{m} \times 0.1 \mu \text{m}$. In the center of the cell, a 20 nm thick gold structure consisting of a triangular antenna and a thin rectangular collector was situated on a 50 nm thick $\text{SiO}_2$ substrate layer, surrounded by air from all other directions, as shown in \cref{fig:simulation-layout}. The triangular antenna had height of 200 nm and base width of 100 nm. The tip of the triangle was rounded with a 10 nm radius of curvature. The rectangular collector had a length of 200 nm and width of 50 nm. The gap between the tip of the triangle and the collector was 40 nm. 

The FDTD simulations of the structure were performed with three sets of boundary conditions: (1) periodic boundaries, (2) perfectly matched layer (PML) boundaries, and (3) absorbing boundaries. Periodic boundaries are implemented as the default boundary in the simulation suite used. PML and absorbing boundaries are available as finite thickness layers that pad a face of the simulation cell. In the simulations with periodic boundaries, 100 nm thick PML layers were used on the top and bottom faces of the simulation cell in the z-direction, with the layer of $\text{SiO}_2$ extending into the PML to prevent reflections which will cause multiple passes of the excitation pulse. In the FDTD simulations with full PML boundaries, 100 nm thick PMLs surrounded the simulation cell from all directions. Likewise, in simulations with full absorbing absorbing boundaries, 100 nm thick absorbing boundaries surrounding the simulation cell in all six directions. 

The incident optical pulse is defined indirectly by defining a current source term located at the top of the simulation cell, directly below the top PML/absorbing layer in the z-direction. The current source term in the simulations was defined as spatially rectangular covering the cross section of the simulation cell with Gaussian pulse temporal characteristics. The resulting incident has a plane-wave front with Gaussian pulse time dependence, but unknown field profile. To calculate the field profile of the optical pulse incident on the structure, we ran an FDTD background simulation in a cell with the $\text{SiO}_2$ layer and the appropriate boundaries, but with no gold structure. 

The spatial discretization of the simulations was set to 250 divisions per $1 \mu \text{m}$. The structure in \cref{fig:simulation-layout} was simulated with incident pulses of different central frequencies and different durations.The time sampling was varied appropriately for each simulation, with samples recorded at the nearest internal time step to integer multiples of $t_{\text{samp}} = \frac{1}{(f_\text{cen}+\frac{1}{2}df)r_{\text{samp}}}$, where $f_{\text{cen}}$ is the central frequency of the source, $df$ is the frequency domain width of the source, and $r_{\text{samp}}$ is a sampling rate. Since the output was not uniformly sampled in time, we added a post processing step in which the fields are interpolated and re-sampled uniformly in time. 

To calculate the transfer function, a fast Fourier transform (FFT) was applied to the results of each of the empty simulations and each of the full structure simulations. With $E_\text{tip}(\omega)$ being the frequency dependent field at the tip of the antenna calculated from the full structure simulation, and $E_\text{inc}(\omega)$ being the frequency dependent incident field calculated by the empty simulation, the transfer function is $H(\omega) = \frac{E_\text{tip}(\omega)}{E_\text{inc}(\omega)}$. The transfer functions from the different frequency bands were stitched together to span the full full frequency range of interest for each boundary type.

\begin{figure}[htbp]
    \centering
    \includegraphics[width=0.98\linewidth]{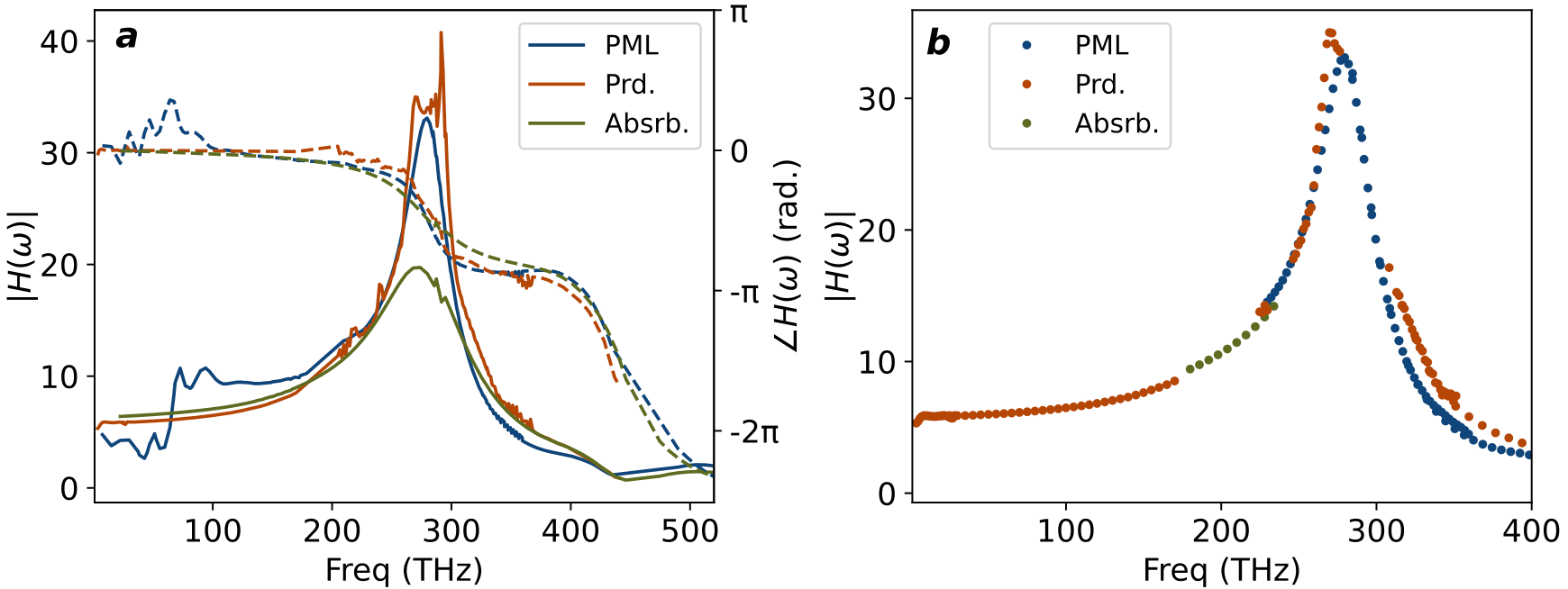}
    \caption{Transfer functions of triangular antennas simulated with PML boundaries, periodic boundaries, and absorbing boundaries. \textbf{a} Transfer function of structure in \cref{fig:simulation-layout} with various boundaries. The full frequency range is achieved from stitching together simulations from different frequency bands. \textbf{b} Subset of simulation data used for the circuit model parameter fitting. In eac frequency band, the data from the simulation with the boundaries which do not have artifacts at that frequency.}
    \label{fig:unit-cell-simulations-transfer-functions}
\end{figure}

 We performed a set of simulations spanning the full frequency domain for each type of boundary: PML, periodic, and absorbing. The full frequency range transfer function for each boundary type is shown in  \cref{fig:unit-cell-simulations-transfer-functions}.\textbf{a}. In simulation cells that are sufficiently small, the boundaries effect the simulated result. Larger cell sizes can avoid boundary artifacts, but they require more memory and more computation time to run. We observe that simulations with PML boundaries have a sudden cliff drop in the low frequencies, which is not known from experiment or present in the periodic simulations. Periodic boundaries inherently simulate an infinite repeating periodic array of the structure with array spacing equal to the cell size. As a result, the transfer function from a periodic simulation has sharp resonance peaks at frequencies associated with the array spacing. Absorbing boundaries clamp the field value at the boundary, and as a result, they diminish the Q-factor of the resonance. For each frequency band, we selected the simulation data that was associated with individual antenna dynamics rather than array resonances or boundary artifacts. The selected data for the circuit model parameter fit is shown in \cref{fig:unit-cell-simulations-transfer-functions}b. 

\subsection{Initial Circuit Parameter Value Calculations} \label{sec:initial-circuit-param-extraction}
The Drude-Lorentz model of free electron gas in plasmas gives an optical conductivity as follows:
\begin{equation}
\sigma(\omega) = \frac{\sigma_0}{1-i\omega \tau} = \frac{\sigma}{1+\omega^2\tau^2} +i \omega \tau \frac{\sigma_0}{1+\omega^2 \tau^2},
\end{equation}
with $\sigma_0 = \frac{N e^2 \tau}{m_e}$. Where $N$ is the free electron density and $m_e$ is the effective mass. In gold, $N=5.9 \times 10^{28} \ \text{m}^{-3}$ \cite{alma990017320150106761} and $m_e = 0.99$ mass of a free electron \cite{PhysRevLett.77.1889}. The impedivity is calculated as $z = \frac{1}{\sigma(\omega)}$. 
\begin{equation}
    z(\omega) = \frac{1 - i \omega \tau}{\sigma_0} = \frac{m_e}{N e^2 \tau} - i \omega \frac{m_e}{N e^2}.
\end{equation}
Real part of the impedivity is the ohmic resistivity, which is independent of frequency. The imaginary part of impedivity is the kinetic inductivity term. The ohmic resistance and the kinetic inductance of a structure can be calculated by multiplying by a length, $l$, and a cross sectional area, $A$.
\begin{equation}
    R = \frac{m_e}{A N e^2 \tau} \frac{l}{A}, \ \ L_k = \frac{m_e}{ N e^2} \frac{l}{A}.
\end{equation}
The value of $\frac{m_e}{A N e^2 \tau}$ is known as the DC resistivity and is $\rho = 2.214 \times 10^{-8} \ \Omega$m for gold at room temperature \cite{matula1979electrical}. With $l=200$ nm and $A= 50 \ \text{nm} \times 20 \ \text{nm}$, $ R = 4.428 \Omega$. $L=0.119$ pH. To improve the initial value of $R$ and $L$ modifications to the Drude model under plasmonic conditions can be considered \cite{moeferdt_plasmonic_2018,kheirandish_modified_2020}
Magnetic inductance is related to the length of the object and the electromagnetic permeability. 
\begin{equation}
    L_m \sim l \mu = 0.251 \text{ pH}
\end{equation}

parallel plate approximation for $C_\text{gap}$ assuming $d=40$ nm: 
\begin{equation}
    C = \frac{\epsilon_0 A}{d} = 0.2214 \text{ aF}
\end{equation}

\subsection{Circuit Parameter Fitting with LOS Optimization} \label{sec:los-fitting}
Prior to optimization, we scaled the circuit variables such that the initial values of the all parameters were within an order of magnitude from one another. The scaling set the basic unit of capacitance to aF, of inductance to pH, and of resistance to $\Omega$. The scaling mitigates prematurely reaching the convergence condition of $|x_i|/||x|| <\text{tol}$ for variables that are orders of magnitude smaller than other variables. 
We define a root mean squared error (RMSE) as a metric for the difference between the scaled analytical transfer function and the FDTD transfer function evaluated at discrete radial frequency sample points $\omega_n$.
\begin{equation} \label{eq:app-rmse-def}
    \text{RMSE} = \frac{1}{\sqrt{N}} \sqrt{\sum_n \big|g H_{\text{ckt}}(\omega_n) - H_{\text{FDTD}}(\omega_n)\big|^2},
\end{equation}
where $n$ is the index of the radial frequency sample points, $N$ is the number of sample points, $H_{\text{ckt}}(\omega_n)$ is the analytical circuit transfer function evaluated at radial frequency $\omega_n$, $g$ is a scaling factor, and $H_{\text{FDTD}}(\omega_n)$ is the stitched transfer function calculated by the FDTD simulations (shown in \cref{fig:unit-cell-simulations-transfer-functions}.\textbf{b}) evaluated at radial frequency $\omega_n$. Since both $H_{\text{ckt}}$ and $H_{\text{FDTD}}$ are complex valued quantities, the definition of the RMSE includes squaring of the magnitude of the difference rather than a direct square of the difference. The scaling factor is obtained by matching the peaks of the circuit transfer function and of the FDTD simulated transfer function such that $g= \frac{|H_{\text{FDTD}}(\omega)|_\text{max}}{|H_{\text{ckt}}(\omega)|_\text{max}}$. The scaling factor is interpreted as the geometrical field enhancement factor at the antenna tip. 

We used a least of squares optimization to minimize the RMSE defined in eq \eqref{eq:app-rmse-def}. The choice of metric and optimization  method gives heavier weight to sections of the spectrum that are more densely sampled and prioritizes matching the resonance peak and quality factor. To evaluate the quality of circuit model fit to the FDTD simulations, we use the $R^2$ metric which is defined as
\begin{equation}
    R^2  = 1 - \frac{\text{RMSE}^2}{\text{Var}(H_{\text{FDTD}}(\omega_n))} = 1- \frac{\frac{1}{N}\sum_{n} |g H_{\text{ckt}}(\omega_n) - H_\text{FDTD}(\omega_n)|^2}{\frac{1}{N}\sum_{n} |H_\text{FDTD}(\omega_n) - \frac{1}{N} \sum_{n} H_\text{FDTD}(\omega_n)|^2},
\end{equation}
with the variance of the complex valued $H_\text{FDTD}(\omega)$ evaluating to a non-negative real number calculated as the sum of the real and imaginary variances. 

\subsection{Wire Simulations} \label{sec:app-tline-simulation}
To simulate the propagation dynamics of a mode on a wire, we produced the closest numerically stable approximation to a simulation of an infinitely long wire with a local line source at its center. Direct simulations of long wires with local sources resulted in non-decaying numerical oscillations due to interactions of boundary effects with the local source. We combined the results of two simulations performed with plane wave sources. In the first simulation, a pair of 2.4 $\mu$m long and 50 nm wide gold wires were spaced at a distance of 115 nm. A 200 nm long triangular antenna was merged with one of the wires such that the tip of the antenna was at a 40 nm distance from the second wire as shown in \cref{fig:transmission-line-char}a. The simulation cell size was 800 nm $\times$ 2.8 $\mu$m $\times$ 100 nm with 100 nm thick PMLs in every one of the six directions surrounding the cell. The incident optical field was polarized along the axis of the antenna, had a plane wave front and was generated by Gaussian source with a central wavelength resonant with the antenna.  The second simulation consisted of an identical pair of 2.4 $\mu$m long and 50 nm wide gold wires were spaced at a distance of 115 nm, but with no antenna (solid lines in \cref{fig:transmission-line-char}a). The optical source in the second simulation was identical to the first one. We subtracted the resulting time dependent near fields on the wires of simulation 2 from the time dependent near fields from simulation 1. The superposition of the simulations effectively isolates the localized mode injected onto the wires by the oscillating antenna resonance response by subtracting out the the distributed wire response to the incident field. 

To capture wire characteristics independent of reflections, we sliced the time domain data and only used the data corresponding to time samples before the arrival of reflections. 

The characteristic impedance of the transmission line, $Z_0$ is obtained by calculating 
\begin{equation}
    Z_0 = \frac{E(\omega)}{H(\omega)},
\end{equation}
where $E(\omega)$ and $H(\omega)$ are the frequency domain electric field and the frequency domain magnetic field respectfully. While $Z_0(\omega)$ is frequency dependent, it is well approximated as a constant over the frequency range of interest. in The phase velocity, $v_p$ is obtained by observing the propagation of a mode on a wire,  and the loss, $\alpha$, was obtained by tracking the loss incurred to a mode while propagating. To then solve for $R,L,C$ for the circuit description of the transmission line, we use the low loss approximations to set initial values $L_i,R_i,C_i$ from the simplified set of equations. 

\begin{equation}
    Z_0   \approx \sqrt{\frac{L_i}{C}_i}
\end{equation}
\begin{equation}
    v_p \approx \frac{1}{\sqrt{L_iC_i}}
\end{equation}
\begin{equation}
    \alpha \approx \frac{R_iC_i}{2}
\end{equation}

With the initial values, we can more easily solve the more complicated system of equations: 
\begin{equation}
    Z_0 = \sqrt{\frac{R+ j \omega L}{j \omega C}},
\end{equation}
The complex $k$ wave vector encodes information about $v_p$ and $\alpha$.  
\begin{equation}
    k = k' + jk'' = \omega \sqrt{LC - j \frac{RC}{\omega}}
\end{equation}

Taking the real part of this, $k'$ to calculate $v_p = \omega/k'$, 

\begin{equation}
    k' = \omega \sqrt{\frac{\sqrt{L^2 C^2 + \frac{R^2 C^2}{\omega^2}} + LC}{2}}
\end{equation} 
\begin{equation}
    v_p = \frac{\sqrt{2}}{\sqrt{\sqrt{L^2C^2+R^2C^2/\omega^2}+LC}},
\end{equation}
\begin{equation}
    \alpha = \sqrt{\frac{\omega^2 \sqrt{L^2C^2 + \frac{R^2 C^2}{\omega^2}} -\omega^2LC}{2}},
\end{equation}

The of confinement of a plasmon to the surface depends of a nanowire depends on the thickness and width of the nanowire. The confinement directly affects the effective wavelength and the phase velocity of the mode on the wire. For that reason, the spatial discretization of the nanowire is a significant factor in simulations used to extract wire characteristics. Since MEEP does not yet incorporate adaptive meshing based on structure, special care should be taken to align the nanowires with the pixel grid. A single pixel offset can cause a 5\%-10\% change in simulated wire width, which could change the characteristics of the wire by a similar factor.  

\subsection{Load Impedance Calculations}
We can identify the load impedance of the wire termination by calculating the reflection coefficient at the load, $\Gamma_L =\frac{Z_L - Z_0}{Z_L + Z_0}$. The reflection coefficient is the ratio of the backwards propagating (reflected) mode and the forward propagating mode. Mathematically, $\Gamma_L = \frac{E^-}{E^+}$. The forward and backward propagating modes can be calculated from the MEEP simulation data in one of two ways.  

Using magnetic field information, we can a simple system of equations and solve for the mode coefficients at the load end of the wires.
\begin{equation}
  E(\omega, 0) =  E^+(\omega, 0) + E^-(\omega, 0) 
\end{equation}
\begin{equation}
    H(\omega, 0) = \frac{1}{z_0} E^+(\omega, 0) - \frac{1}{z_0} E^-(\omega, 0), 
\end{equation}
where the $y=0$ point is at the load end of the wire. Having calculated the characteristic impedance of the wire previously, we can solve for the forward and backwards mode coefficients. Each point in frequency requires a system of equation solve. Using the MEEP simulation data, we solved for the reflection coefficient, $\Gamma_L$ over a range of frequency and used that to calculate the impedance of the load. 

\begin{figure}[htbp]
    \centering
    \includegraphics[width=0.7\linewidth]{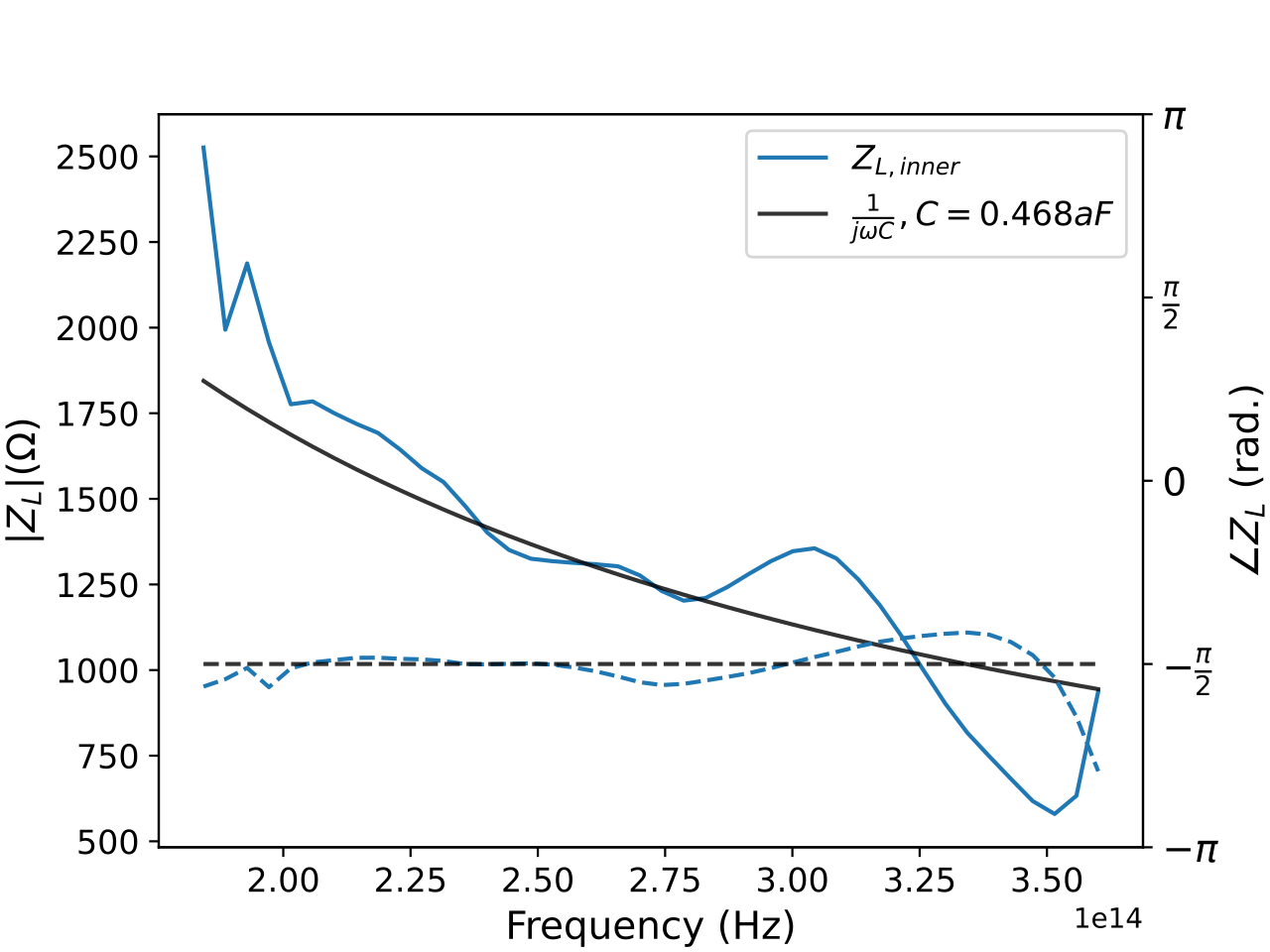}
    \caption{Impedance of load at the end of floating wires connected to an antenna source. The blue trend is from numerical simulation data. The black curve is a capacitor fit.}
    \label{fig:load-impedance-0calculation}
\end{figure}

The calculated load impedance attained from the MEEP simulation data is capacitive. We fit the load impedance to a capacitor of $0.468$ aF. 

\subsection{LTspice Simulations Unit Transform} \label{sec:spice-transf}
The LTspice simulation seems to have a minimum time step which isn't easily identifiable. To be able to perform simulations with femtosecond dynamics, we introduce a unit transformation that transforms a femtosecond in real time to a second in an LTspice simulation. The full transformation is:
We choose a scaling factor $\tau$ to control the time unit scaling, such that 
\begin{equation}
    t_n = \tau t,
\end{equation}
where $t_n$ is the simulation time unit and $t$ is the real-life time unit. We seek to conserve charge, so $Q = Q_n$. We let that control the time constant for current, $I$,
\begin{equation}
    I_n = \frac{dQ_n}{dt_n} = \frac{dQ}{\tau dt} = \frac{1}{\tau} I.
\end{equation}
Next, we look at the capacitance equation $i = C\frac{dv}{dt}$, we choose a scaling $T$ to account for capacitance scaling, $C_n = TC$, that sets the voltage ratio from the capacitance equation
\begin{equation}
    I_n = C_n \frac{dV_n}{dt_n} = I\tau^{-1} = CT \frac{dV_n}{\tau dt}.
\end{equation}
The $\tau$ factors cancel out and 
\begin{equation}
    I = CT \frac{dV_n}{dt} \ \ \rightarrow \ \ V_n = T^{-1} V.
\end{equation}
Next, we use Ohm's law to set the resistance, $R_n$
\begin{equation}
    V_n = I_n R_n = T^{-1} V = \tau^{-1} I R_n \ \ \rightarrow \ \ R_n = \frac{T^{-1}}{\tau^{-1}}\frac{V}{I} = \frac{\tau}{T} R.
\end{equation}
Next, we set the inductance, 
\begin{equation}
    V_n = L_n \frac{dI_n}{dt_n} = T^{-1}V = L_n \frac{\tau^{-1} dI}{\tau dt} \ \ \rightarrow \ \ L_n = V \left(
\frac{dI}{dt} \right)^{-1} \frac{\tau^2}{T} = \frac{\tau^2}{T} L.
\end{equation}
We set $\tau = 10^{15}$ and $T = 10^{12}$, leading to the unit conversion in\cref{tab:spice-unit-transf}. 

\subsubsection{Unit Transformation Consistency Check:}
\textbf{Energy for inductor: }
\begin{equation*}
  \frac{1}{2} L_n I_n^2 = \frac{1}{2} (T^{-1}\tau^2) L ( \tau^{-1} I)^2 = \frac{1}{2} LI^2 T^{-1} = \frac{1}{2} T^{-1} J 
\end{equation*}
\textbf{Energy for capacitor: }
\begin{equation*}
   \frac{1}{2} C_n V_n^2 = \frac{1}{2} T C (T^{-1} V)^2 =  \frac{1}{2} T^{-1} J 
\end{equation*}
\textbf{RC and LC time constants:}
\begin{equation*}
    R_nC_n = \frac{\tau}{T} R \times T C= \tau s, \ \ \frac{L_n}{R_n} = \frac{\tau^2 L}{T} \frac{T}{\tau R} = \tau s
\end{equation*}
\textbf{Resonance and Q factor for series RLC:}
\begin{equation*}
 \omega_{0,n} = \frac{1}{\sqrt{L_n C_n}} = \frac{1}{\sqrt{\frac{\tau^2}{T}L TC}} = \frac{1}{\sqrt{\tau^2}} = \tau^{-1} s
 \end{equation*}
 \begin{equation*}
Q_n = \frac{1}{R_n}\sqrt{\frac{L_n}{C_n}} = \frac{T}{\tau} \sqrt{\frac{T^{-1} \tau^2 L}{TC}} = \frac{T}{\tau}\sqrt{\frac{\tau^2}{T^2}} = 1,
\end{equation*}
where $Q_n$ is the quality factor, not the charge. \\
\textbf{Resonance and Q factor for parallel RLC:} \\
Resonance frequency is the same as parallel combination, but the quality factor is given as 
\begin{equation*}
    Q_n = R_n \sqrt{\frac{C_n}{L_n}} =1
\end{equation*}
because this is the inverse of the quality factor of the series combination. 

\begin{table}[htbp]
    \centering
    \begin{tabular}{|lccc|}
    Quantity & Conversion relation & Spice basic unit & Equivalent SI unit\\
    \hline
       Capacitance: &  $C_n = TC$& $F_n$ & $10^{-12} F$ \\
Inductance:  & $L = \tau^2 T^{-1} L$ & $H_n$ & $10^{-18} H$ \\
Resistance: & $R_n = T\tau ^{-1} R$ &$\Omega_n$ & $10^{-3} \Omega$ \\
Time: & $t_n = \tau t$ &$s_n$ & $10^{-15} s$ \\
Voltage: &$V_n = T^{-1}V$ & $V_n$ & $10^{12} V$ \\
Current:  & $I_n = \tau^{-1} I$& $A_n$ & $10^{15} A$ \\ 
Frequency & $\omega_n = \tau \omega$ &$\omega_n$ & $10^{15} Hz$
    \end{tabular}
    \caption{LTspice unit conversion transformation. Table values assume $T=10^{12}$ and $\tau = 10^{15}$.}
    \label{tab:spice-unit-transf}
\end{table}

It should be noted that due to the scaling in voltage and current, special care needs to be taken with non-linear elements. In the voltage controlled current sources defined by us, we transform voltages back to SI units before calculating the corresponding fields and plugging the value into the emission rate equation. The tunneling-emission current is scaled by the corresponding $10^{-15}$ factor before it is injected into the circuit. Standard non-linear elements from LTspice library will not work in this environment without user modification to account for the scaling. 

In LTspice there are two main models for transmission lines: lossless and lossy transmission lines. The lossless transmission line model is simpler and therefore, generally computationally less costly than the lossy transmission line model for comparable transmission line lengths. For relatively short wires, a lossy transmission line can be approximated by a lossless transmission line in series with a resistor. In the LTspice implementation of the model, we compared the accuracy of lossless and lossy transmission lines. Ultimately, we continued with lossy transmission lines for the more complex circuits out of a desire to stick to maximally accurate models in the compact model. 

\subsection{Coupling of Antenna to Wires} \label{sec:delta-phi-ttl-deriv}
We performed a set of time domain simulations to explore the dynamics of an antenna coupled into a pair of wires. The antenna was specified as a triangle with 200 nm height and 100 nm base, and the pair of wires were each 50 nm wide and were spaced 115 nm apart. The lengths of the wires were varied between 70 to 300 nm throughout the simulation set. One wire was merged with the collector, and the other wire intersected the triangle at the center as shown in \cref{fig:wire-char-fdtd-sim}a. Each antenna-wire system was simulated with three different input pulse frequencies, to gather information about the full frequency range of interest. MEEP was used as the simulator and each simulation was performed with PML boundaries and periodic boundaries.

The transfer function of each antenna-wire system was extracted from each simulation for its corresponding frequency band. Then, the full frequency range transfer function for each antenna-wire system was calculated by stitching together the transfer functions of the various frequency bands. An analysis of the transfer functions reveals a resonant dip feature which digs into the resonance of the antenna. As the length of the wires increases, the resonant dip shifts towards the lower frequencies. The dip in the resonance of the transfer function is likely due to destructive interference between modes at the tip of the antenna due to reflections from the edge of the wire. To interfere, a mode must complete a round trip from the source, to the load, and back to the source accumulating a phase shift along the travel path and from the reflection. Additional phase is accumulated from the travel along the antenna surface to get to the wire. 
\begin{equation} 
    \Delta \phi_{\text{ttl}} = \Delta \phi_\text{travel} + \Delta \phi_\Gamma + \Delta \phi_{\text{tip}} = m \pi, \ m \in \{1, 2, 3, 5, \dots\} \label{eq:delta_phi_ttl_appdx}
\end{equation}
For destructive interference, the total accumulated phase must be an odd multiple of $\pi$, or $\Delta \phi_{\text{ttl}} = m\pi$ where $m = \{1, 3, 5, \dots\}$. The tip of the antenna is modeled as the source. 

The phase shift accumulated by a mode over a travel distance of $L$ is 
\begin{equation}
    \Delta \phi_\text{travel} = \frac{2 \pi L}{\lambda } = \beta L,
\end{equation}
where $\beta$ is the propagation mode $\beta = \frac{2 \pi}{\lambda}$. Here, $\lambda$ is the wavelength of a confined surface mode, also written as $\lambda = \frac{\lambda_0}{n}$ where $n$ is the index of refraction and $\lambda_0$ is the vacuum wavelength of the mode. The surface confined mode can also be written as $\lambda = \lambda_0 \frac{v_p}{c}$ with $v_p$ as the phase velocity.

The phase shift due to reflection at the metal-air interface at the end of the wire is given by the reflection coefficient, $\Gamma$. We determined that $Z_L = \frac{-j}{\omega C}$ is a capacitive load of 0.468 aF. Therefore, 
\begin{equation}
    \Gamma = \frac{Z_L - Z_0}{Z_L + Z_0} = \frac{-\frac{j}{\omega C} -Z_0 }{-\frac{j}{\omega C} + Z_0}
\end{equation}
The total phase accumulation is 
\begin{equation}
    \Delta \phi_{ttl} = 2 \frac{2 \pi L n}{\lambda_0} + \Delta \phi_\Gamma + \Delta \phi_{tip} = m \pi, \ m \in \{1, 2, 3, 5, \dots\}
\end{equation}

The load impedance at the end of the wire takes the form of a capacitive load. That allows us to calculate the phase of the reflected mode. 

\begin{equation}
    \Gamma =  \frac{-\frac{j}{\omega C} -Z_0 }{-\frac{j}{\omega C} + Z_0} \frac{\frac{j}{\omega C} +Z_0}{\frac{j}{\omega C} +Z_0} = \frac{-(\frac{-1}{\omega^2 C^2} +Z_0^2 +\frac{2 j Z_0}{\omega C})}{Z_0^2 + \frac{1}{\omega^2 C^2 }}
\end{equation}
With $\omega = 2 \pi f = \frac{2 \pi c}{\lambda_0}$, $c$ is the speed of light. 

\begin{equation}
    \Gamma = \frac{\frac{1}{(2\pi c/\lambda_0)^2C^2} - Z_0^2 -\frac{2 j Z_0}{2\pi c C/\lambda_0}}{Z_0^2 + \frac{1}{(2\pi c/\lambda_0)^2 C^2}} = \frac{\frac{\lambda_0^2}{4 \pi^2 c^2 C^2} -Z_0^2 -\frac{2j Z_0 \lambda_0}{2 \pi c C}}{Z_0^2 + \frac{\lambda_0^2}{4 \pi^2 c^2 C^2}}
\end{equation}
The Phase of $\Gamma$ is 
\begin{equation}
    \Delta \phi_\Gamma = \tan^{-1} \left( \frac{-2Z_0 \lambda_0/(2\pi c C )}{\lambda_0^2/(4 \pi^2 C^2 c^2) -Z_0^2}\right) = \tan^{-1} \left( \frac{-4 \pi Z_0 \lambda_0 C c}{\lambda_0^2 - 4\pi^2 C^2 c^2 Z_0^2}\right)
\end{equation}
We know that $C$ is very small, roughly on the order of an aF or below. $\lambda_0$ is on the order of a micron, $Z_0$ is roughly $500 \Omega$. We can take the small x approximation of $\tan^{-1}(x)$. 
\begin{equation}
    \Delta \phi_\Gamma \approx \frac{-Z_0 \lambda_0 4 \pi C c}{\lambda_0^2 - 4\pi^2 C^2 c^2 Z_0^2} + \frac{1}{3} \left( \frac{Z_0 \lambda_0 4 \pi C c}{\lambda_0^2 - 4\pi^2 C^2 c^2 Z_0^2}\right)^3 - \frac{1}{5} \left( \frac{Z_0 \lambda_0 4 \pi C c}{\lambda_0^2 - 4\pi^2 C^2 c^2 Z_0^2}\right)^5  \dots
\end{equation}
We can take the first term only because $C$ scales the quantity to be very small. We now have a more detailed expression for the phase accumulated 
\begin{equation}
    \Delta \phi_{ttl} = \frac{4 \pi L n}{\lambda_0} - \frac{Z_0 \lambda_0 4 \pi C c}{\lambda_0^2 - 4\pi^2 C^2 c^2 Z_0^2} + \Delta \phi_{tip} = m \pi, \ \ m \in \{1, 3, 5, \dots\}
\end{equation}
 Let's assume that the $Z_0^2 C^2 c^2$ term in the denominator is negligible to $\lambda_0^2$. 
 \begin{equation}
     \frac{4 \pi L n}{\lambda_0} - \frac{Z_0 4 \pi C c}{\lambda_0} = m\pi - \Delta \phi_{tip} 
 \end{equation}
 \begin{equation}
     \lambda_0 = \frac{4\pi(Ln - Z_0 C c)}{m \pi - \Delta \phi_{tip}}
 \end{equation}

\subsection{Optical Effects of Electrons in Transit in Vacuum} \label{sec:append-optical-effects-of-plasma}
The optical effects of the electron plasma can be expressed as a kinetic inductance term, which can be calculated with the Drude model assuming a uniform electron plasma in the gap.
\subsubsection{Direct Calculation Assuming Uniform Electron Plasma} \label{sec:app-direct-calc-lk}
The total kinetic energy of a collection of electrons is 
\begin{equation} \label{eq:kinetic-e}
    \varepsilon_k = \frac{1}{2} \sum_q m_{e} v_{e,q}^2
\end{equation}
where $q$ is the summation index of the electrons, $m_e = 9.11\times 10^{-31}$ Kg is the mass of an electron, and $v_{e,q}$ is the velocity of electron $q$. The directed motion of charge carriers can be defined as a current such that 
\begin{equation}
    i = \sum_q e v_{e,q} = A n e v, 
\end{equation}
where $i$ is the current, $e = 1.602 \times 10^{-19}$ C is the charge of the electron. When individual electron properties are unknown, the current can be defined based on aggregate statistics. $A$ is the cross sectional area of the electron flux, $n_e$ is the three-dimensional electron density in the vacuum channel, and $v$ is the average electron velocity. Substituting $v = i/(A N e)$ into $ v_{e,q}$ eq. \ref{eq:kinetic-e}, yields
\begin{equation}
    \varepsilon_k = \frac{1}{2} \sum_q m_e \frac{i^2}{A^2 N^2
e^2}.
\end{equation}
The number of electrons in the gap is equal to $A \times \ell \times n_e$, where $\ell$ is the length of the gap. The final expression for the kinetic energy based on the current definition, is:
\begin{equation}
\varepsilon_k = \frac{1}{2} \frac{m_e \ell}{A n e^2} i^2.
\end{equation}
From the resemblance to Faraday inductance in magnetic potential energy, $\varepsilon_F = \frac{1}{2} L_F I^2$, we take the analogous quantity as the kinetic inductance 
\begin{equation}
    L_k = \frac{m_e \ell}{A n_e e^2} = \frac{m_e \ell^2}{\# \text{electrons} \ e^2}
\end{equation}

We followed experimental results from Ritzkowsky et al. for numerical estimates of the kinetic inductance \cite{ritzkowsky_-chip_2024}. The authors describe a peak electron current of 92 electrons/fs over for a pulse with a half-cycle duration of 4.5 fs. The effective emitting surface area is $A = 628 \text{nm}^2$, and the gap distance is $\ell = 50$ nm. We estimate the electron density as $n_e = \frac{92}{628 \text{nm}^2 \times 50 \text{nm}} = 2.9299 \times 10^{24} \ \text{m}^{-3}$.
The resulting kinetic inductance is:
\begin{equation} \label{eq:ch5-lk-current}
    L_k = \frac{m_e \ell}{A n e^2} = \frac{9.11 \times 10^{-31} \ \text{Kg} \cdot 50 \ \text{nm}}{628 \ \text{nm}^2 \cdot 2.9299 \times10^{24} \ \text{m}^{-3} \cdot (1.602 \times 10^{-19} \text{C})^2} = 9.6461 \times 10^{-10} \ \text{H}.
\end{equation}
The vacuum gap kinetic inductance estimate of $\sim 1$ nH is four orders of magnitude larger than the kinetic inductance of the antenna emitter. 

The kinetic inductance estimate above is based on the assumption that electrons are distributed as a uniform cloud in the gap of nanostructures. Since that assumption is not directly verifiable, we introduce a second analysis based on known and verifiable quantities. 

\subsubsection{Kinetic Inductance from Measurable Peak Current} \label{sec:app-current-calc-lk}
We know that electrons are tunneled through the barrier with zero kinetic energy and that they gain momentum from the fields and the ponderomotive potential. Therefore, the kinetic energy of the electrons traversing the gap is: $U_p N_e$ where $U_p$ is the ponderomotive potential and $N_e$ is the number of electrons. We can set the total kinetic energy of the moving electrons to the inductive energy of the current. 
\begin{equation}
    \frac{1}{2} L_k I_p^2 = U_p N_e
\end{equation}
where $I_p$ is the peak current observed. The number of electrons can be rewritten as $N_e = I_p \Delta t/e$ where $\Delta t$ is the emission window. The current $I_p = 14 $ mA and we assume that $\Delta t =1$ fs. With the appropriate substitutions, and assuming $U_p \approx 1 $ eV, we get 
\begin{equation} \label{eq:ch5-lk-energy}
    L_k = 2\frac{e U_{p, J} I_p \Delta t}{ I_p^2e} = 2\frac{1\times 10^{-15} \times 1}{1.4 \times 10^{-2}} = 1.43 \times 10^{-13} H.
\end{equation}

\subsubsection{Optical Simulation of Electron Plasma} \label{sec:app-optical-sim-of-n-in-gap}
\begin{figure}[htbp]
    \centering
    \includegraphics[width=0.8\linewidth]{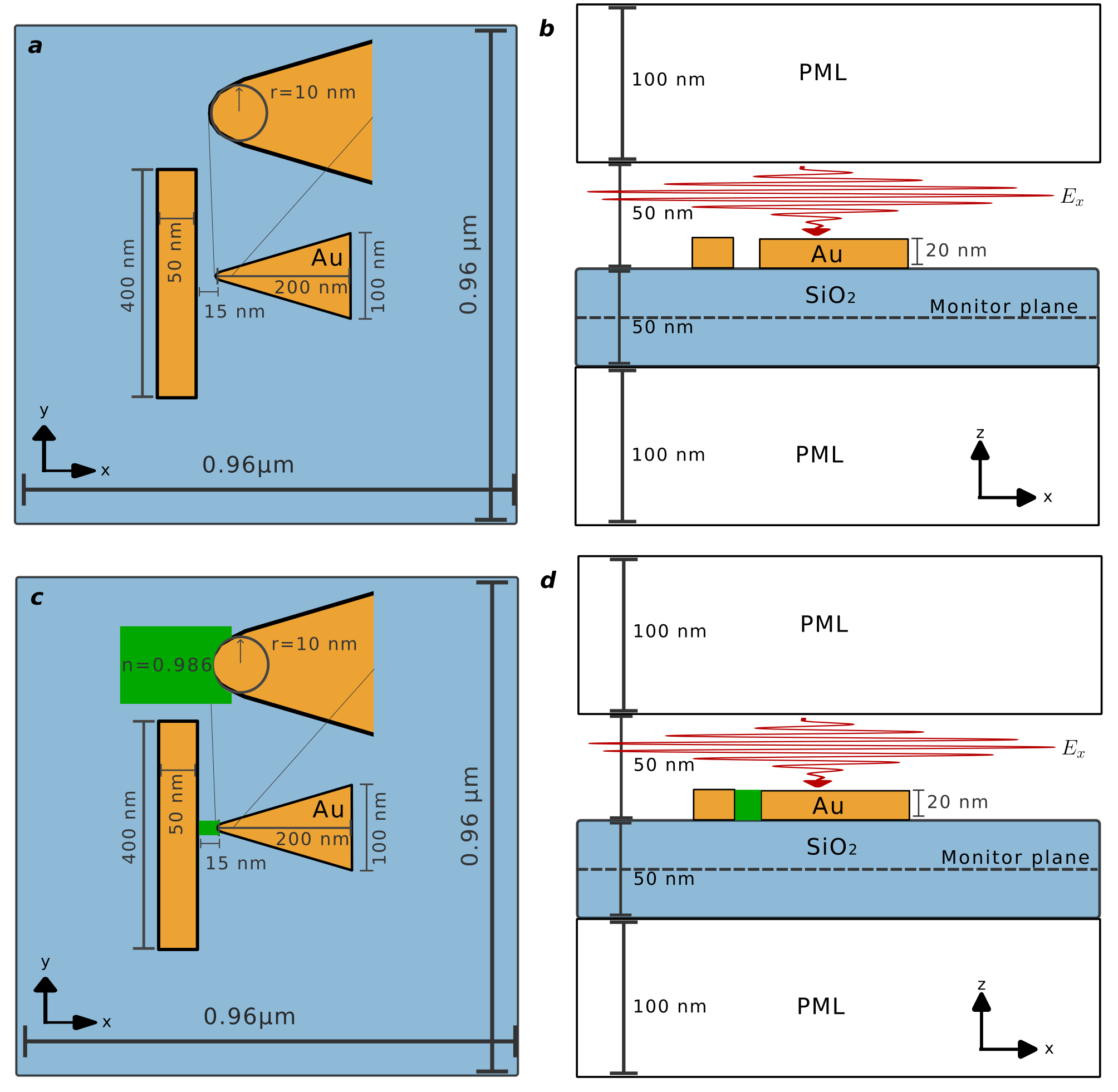}
    \caption{FDTD simulation of optical effects of electrons in gap of nano-emitter. \textbf{a-b} Top-down view and horizontal cross section of a simulation cell with a gold nanoantenna with an incident optical pulse. \textbf{c-d} Top-down view and horizontal cross section of simulation cell featuring a gold nanoantenna with a pseudo material of refractive index $n=0.986$ in the antenna gap with. }
    \label{fig:n-in-gap-sim-cell}
\end{figure}

We consider a collisionless plasma under the Drude model such that the polarization density is given as function of the electric field $P = \epsilon_0 (1 + \epsilon_r)E$. The relative permittivity, $\epsilon_r$ is given by
\begin{equation}
    \epsilon_r (\omega) = 1 - \frac{\omega_p^2}{\omega^2},
\end{equation}
Where $\omega_p$ is the plasma frequency defined as $\omega_p =\sqrt{\frac{n_e e^2}{m_e \epsilon_0}} = 9.6552 \times 10^{13} \ \text{rad}^{-1}$. As calculated in \cref{sec:app-direct-calc-lk}, $n_e = 2.9299 \times 10^{24} \ \text{m}^{-3}$ is the three-dimensional electron density, $e$ is the electron charge, $m_e$ is the electron mass, and $\epsilon_0$ is the permittivity of free-space. We calculated the refractive index of the vacuum channel with the emitted electrons in it. Assuming incident field of $\lambda_0 = 1063$ nm, with radial frequency $\omega = 1.772 \times 10^{15} \ \text{rad}^{-1}$, $n$ is:
\begin{equation} \label{eq:n}
    n = \sqrt{\epsilon_r} = \sqrt{1 - \frac{\omega_p^2}{\omega^2}} = \sqrt{1-\frac{9.3223 \times 10^{27} \ \text{rad}^{-2}}{3.1394 \times 10^{30} \ \text{rad}^{-2} }} = 0.99851.
\end{equation}
To test the effects of the emitted charge in the gap on the antenna's optical properties, we performed a pair of finite difference time domain simulations. As shown in fig \ref{fig:n-in-gap-sim-cell}, in the first simulation a triangular nanoantenna was excited by a polarized pulse. In the second simulation, a pseudo material with refractive index $n=0.986$ was placed in the antenna's gap. It should be noted that $n=0.986$ is smaller refractive index than the one calculated in eq.\ref{eq:n} based on \cite{ritzkowsky_-chip_2024}. This value was selected in order to exaggerate the optical effects.
The simulations were used to extract the transfer function of the antenna for $n=1$ in the gap and $n=0.986$ in the gap. The complex transfer function is calculated as 
\begin{equation}
H(\omega) = \frac{\mathcal{F}\{E_{\text{tip}}\}(\omega)}{\mathcal{F}\{E_{\text{inc}}\}(\omega)},
\end{equation} 
where $E_{\text{tip}}$ is the time dependent field at the tip, $E_{\text{inc}}$ is the time dependent incident field, $\mathcal{F}$ denotes a Fourier transform. A monitor plane in the center of the $\text{SiO}_2$ layer was used to record the electric fields of a traveling mode after interacting with the antenna.

\begin{figure}[htbp]
    \centering
    \includegraphics[width=0.7\linewidth]{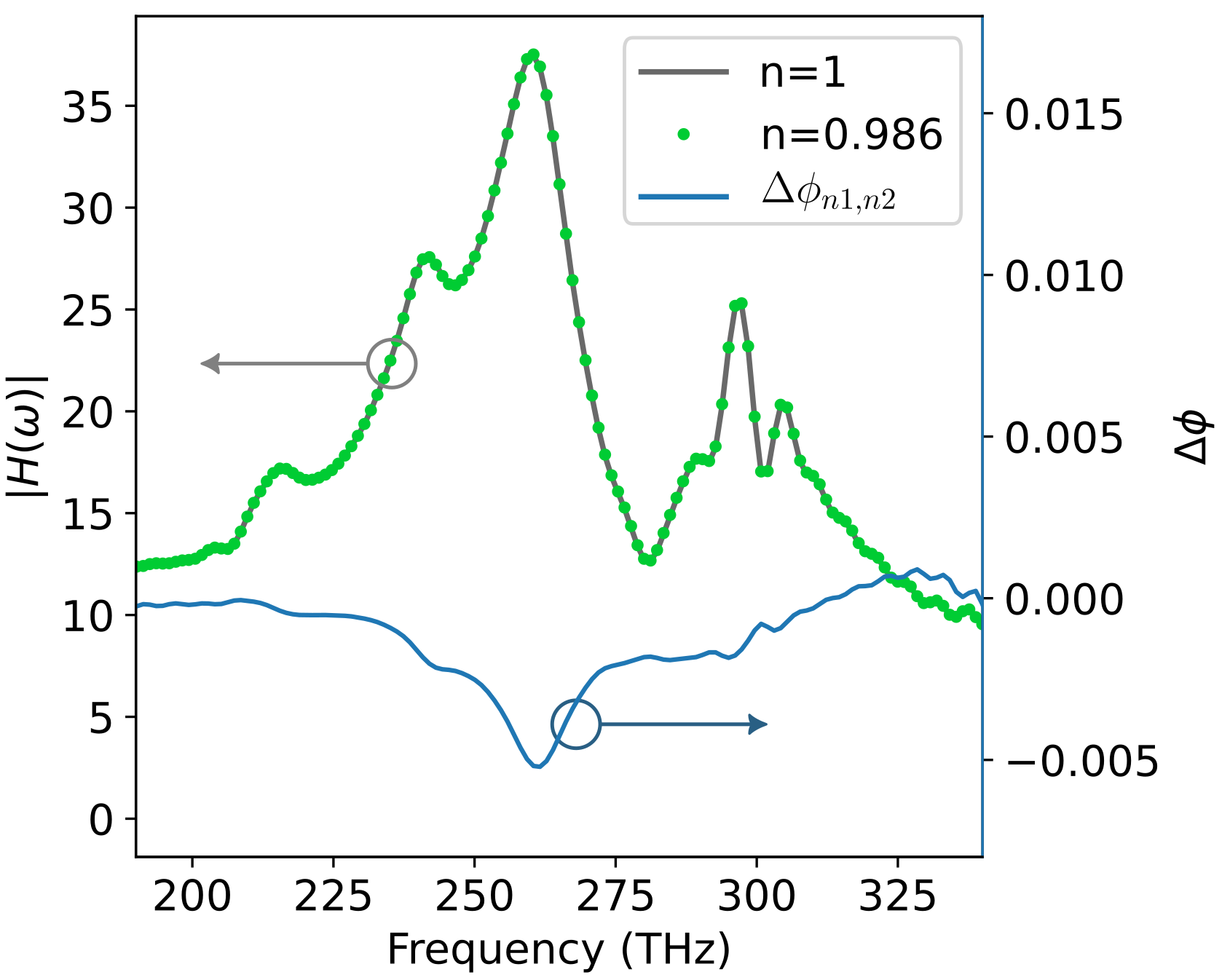}
    \caption{Simulation results. Solid gray line and green dots are the transfer function magnitude of antenna with vacuum gap and of antenna with $n=0.986$ pseudo material in gap respectively. Blue trace shows the phase difference, $\Delta \phi_{n1,n2}$ between the traveling modes in simulation 1 and the traveling modes in simulation 2.}
    \label{fig:n-in-gap-sim-results}
\end{figure}
Figure \ref{fig:n-in-gap-sim-results} shows the results of the simulations. Even with a pseudo material with an exaggerated index, the optical fields on the antenna do not change by a noticeable amount, as seen by the almost perfect overlap of the solid gray trace and the dotted green trace corresponding to the transfer functions of the antennas in both simulations. To get a better metric of the optical effects due to the pseudo material in the gap, we calculate the phase shift incurred by a mode traveling through the pseudo material. The phase shift is calculated as follows
\begin{equation}
    \Delta \phi_{n1, n2} = \angle E_{\text{monitor}, n1} (\omega)- \angle E_{\text{monitor}, n2} (\omega),
\end{equation}
where $\angle E_{\text{monitor}, n1}  (\omega)$ ($\angle E_{\text{monitor}, n2}(\omega)$) is the phase of the frequency dependent electric field sampled on the monitoring plane in the $\text{SiO}_2$ layer in simulation 1 (2) where $n=1$ ($n=0.986$). The phase shift shown on the right y-axis of fig.\ref{fig:n-in-gap-sim-results} and has maximal absolute value of 5 milli-radians. We expect the phase shift to be $\Delta \phi{n1,n2} = \frac{2 \pi}{\lambda} \Delta n L$, where $\Delta n = 0.014$ is the difference between the refractive index of vacuum and of the pseudo material, and $L=20$ nm is the interaction length. 

If we wish to include the kinetic inductance of the vacuum channel in the circuit model, it must be added in series to the current source controlling charge emission. If added in parallel to the current source, regardless if the value from \cref{sec:app-direct-calc-lk} or the value from \cref{sec:app-current-calc-lk} is used, the inductance will obliterate the resonance of the antenna. As demonstrated by the optical simulations in \cref{sec:app-optical-sim-of-n-in-gap} and by the experimental result in \cite{bionta_-chip_2021}, the antenna resonance is preserved even when emission is active and tunnel-emitted electrons flood the gap. We conclude that the electrons on the surface of the metal are shielded by the potential barrier from the electrons in the gap, which preserved the charge oscillations that are inherent to the resonance response. To protect the antenna resonance in the circuit model, we can place the kinetic inductance in series with the current source, which acts as an open circuit when current is not flowing.  